\documentclass[ste,twoside]{stefano}

\usepackage{amsmath}
\usepackage{amssymb}
\usepackage{graphics}
\usepackage{rotating}
\usepackage{cite}

\def\makeheadbox{{%
\hbox to0pt{\vbox{\baselineskip=10dd\hrule\hbox
to\hsize{\vrule\kern3pt\vbox{\kern3pt
\hbox{  {\sf Journal of  Mathematical Physics} {\bf 42}, 2236-2265 (2001)}
\hbox{  {\sf math-ph/0005023} \hspace*{11.3cm} 
$\boldsymbol{\Sigma \delta \Lambda}$ }
\kern3pt}\hfil\kern3pt\vrule}\hrule}%
\hss}}}

\def\fq{\mbox{\tiny $\frac{1}{4}$}}

\def\infm{\mbox{\tiny $- \infty$}}
\def\infp{\mbox{\tiny $+ \infty$}}
\def\+{\mbox{\tiny $\dag$}}
\def\u{\mbox{\tiny $u$}}

\def\a{\mbox{\tiny $a$}}
\def\b{\mbox{\tiny $b$}}

\def\0{\mbox{\tiny $0$}}
\def\1{\mbox{\tiny $1$}}
\def\2{\mbox{\tiny $2$}}
\def\3{\mbox{\tiny $3$}}
\def\4{\mbox{\tiny $4$}}
\def\5{\mbox{\tiny $5$}}
\def\6{\mbox{\tiny $6$}}
\def\7{\mbox{\tiny $7$}}
\def\8{\mbox{\tiny $8$}}
\def\9{\mbox{\tiny $9$}}
\def\i{\mbox{\tiny $i$}}
\def\j{\mbox{\tiny $j$}}
\def\k{\mbox{\tiny $k$}}
\def\m{\mbox{\tiny $m$}}
\def\p{\mbox{\tiny $p$}}
\def\q{\mbox{\tiny $q$}}
\def\r{\mbox{\tiny $r$}}
\def\t{\mbox{\tiny $t$}}
\def\x{\mbox{\tiny $x$}}
\def\xx{\mbox{\tiny $xx$}}
\def\xxxx{\mbox{\tiny $xxxx$}}
\def\y{\mbox{\tiny $y$}}
\def\z{\mbox{\tiny $z$}}
\def\n{\mbox{\tiny $n$}}
\def\sdag{\mbox{\tiny $\dag$}}
\def\p{\mbox{\tiny $p$}}

\def\nxn{\mbox{\small $n \times  n$}}

\def\pem{\mbox{\tiny $\pm$}}
\def\mi{\mbox{\tiny $-$}}
\def\pl{\mbox{\tiny $+$}}
\def\={\mbox{\tiny $=$}}
\def\T{\mbox{\tiny T}}
\def\L{\mbox{\tiny $L$}}
\def\R{\mbox{\tiny $R$}}
\def\Re{\mbox{\tiny $\mathbb{R}$}}
\def\Co{\mbox{\tiny $\mathbb{C}$}}
\def\Ha{\mbox{\tiny $\mathbb{H}$}}

\def\Lo{\mbox{\tiny L}}
\def\Sh{\mbox{\tiny S}}

\begin{document}
%

\title{QUATERNIONIC DIFFERENTIAL OPERATORS}

\author{
Stefano De Leo\inst{1}
\thanks{Partially supported by the FAPESP grant 99/09008--5.}
\and  
Gisele Ducati\inst{1,2}
\thanks{Supported by a CAPES PhD fellowship.}
}

\institute{
Department of Applied Mathematics, University of Campinas\\ 
PO Box 6065, SP 13083-970, Campinas, Brazil\\
{\em deleo@ime.unicamp.br}\\
{\em ducati@ime.unicamp.br}
\and
Department of Mathematics, University of Parana\\
PO Box 19081, PR 81531-970, Curitiba, Brazil\\
{\em ducati@mat.ufpr.br}
}


\date{Accepted for publication in JMP: {\em February 6, 2001}}

\abstract{
Motivated by a  quaternionic formulation of quantum mechanics,
we discuss quaternionic and complex linear differential equations.
We touch only a few aspects of the mathematical theory, namely  
the resolution of  the second order differential equations with constant 
coefficients.  We overcome the  problems coming out from  
the loss of the  fundamental theorem of the algebra for quaternions and 
propose  a practical method to solve quaternionic
and complex linear  second order differential equations with constant 
coefficients.  The resolution of the complex linear Schr\"odinger equation,
 in presence of quaternionic  potentials, represents an interesting 
application of the mathematical material discussed in this paper.}





\maketitle


\section{INTRODUCTION}

There is a substantial literature analyzing the possibility to
discuss quantum systems by adopting quaternionic wave 
functions~\cite{FIN62,FIN63a,FIN63b,ADL88,DAV90,DEL92,ADL94b,ADL94a,DEL95b,DEL96a,DEL96e,ADL96c,DEL98a,DEL98b}. 
This research field has been attacked by a number of people
leading to substantial progress. 
In the last 
years, many articles~\cite{DAV89,DAV92,DEL95a,SCO95,ADL96,DEL96c,DEL96b,ZHA97,SCO97,ADL97,DRA98a,DRA98b,DRA99a,DRA99b,DEL99a,BAK99,DEL00}, review 
papers~\cite{HOR84,ADL94d,DEL97c}  
and books~\cite{DIX94,ADL95a,GUR96} provided a detailed 
investigation  of 
group theory, eigenvalue problem, scattering theory, relativistic wave 
equations, Lagrangian formalism and variational calculus within 
a quaternionic formulation of quantum mechanics and field theory.  
In this context, by observing that the formulation of physical problems in 
mathematical terms  often requires the study of partial differential 
equations, we develop the necessary theory to solve  quaternionic and complex 
linear differential equations.  
The main difficulty in carrying out the solution 
of quaternionic differential equations 
is obviously represented by the non commutative nature of the quaternionic 
field.  The standard methods of resolution break down  and, 
consequently, we need to modify the classical approach. 
It is not our purpose to develop a complete 
quaternionic theory  of differential equations. This exceeds the scope of 
this paper. The main objective is to include what seemed to be 
most important for an introduction to this subject. 
In particular,  we restrict 
ourselves to second order differential equations and  
give a practical method to solve such equations when quaternionic 
constant coefficients appear.

Some of the results given in this paper can be obtained by  
translation into a complex formalism~\cite{DAV89,DAV92,DEL00}.   
Nevertheless, many  subtleties of quaternionic calculus 
are often lost by using  the translation trick. 
See for example, the difference between quaternionic and complex geometry
in quantum mechanics~\cite{HOR84,DEL97c}, generalization of  variational
calculus~\cite{DEL95b,DEL96a}, the choice of a 
one-dimensional quaternionic 
Lorentz group for special relativity~\cite{DEL96b}, the 
new definitions of 
transpose and determinant for quaternionic matrices~\cite{DEL99a}.
A wholly quaternionic derivation of
the general solution of second order differential equations requires
a detailed discussion of the fundamental theorem of algebra for quaternions,
a  revision of the resolution methods and 
a quaternionic generalization of the complex results.

The study  of quaternionic 
linear second order differential equations with constant coefficients 
is based on the explicit resolution of the characteristic quadratic 
equation~\cite{NIV41,NIV42,BRA42,EIL44}. 
We shall show that the loss of
fundamental theorem of the algebra for quaternions does not represent a 
problem in solving quaternionic linear second order differential equations
with constant coefficients.  From there, we introduce more 
advanced concepts, like diagonalization and Jordan form for quaternionic
and complex linear matrix operators, which are developed in detail in the 
recent literature~\cite{ZHA97,SCO97,ADL97,DRA98a,DRA98b,DRA99a,DRA99b,DEL99a,BAK99,DEL00} and we apply them to solve quaternionic and 
complex linear  second order differential equations
with constant coefficients.

As application of the mathematical material presented in this paper, 
we discuss the complex linear Schr\"odinger 
equation in 
presence of quaternionic potentials and solve such an equation for stationary 
states and constant potentials. We also calculate the relation between 
the reflection and transmition coefficients for the step and square potential 
and give the quaternionic solution for bound states.

This work  was intended as 
an attempt at motivating the study of quaternionic and 
complex linear differential equations in view of their future   
applications within a quaternionic formulation of quantum mechanics.
In particular, our future objective is to understand the role  that
such equations could play  in developing  non relativistic quaternionic 
quantum dynamics~\cite{ADL88}  and the meaning that    
quaternionic potentials~\cite{DAV89,DAV92}
could play in discussing  CP violation in the kaon system~\cite{ADL95a}.

In order to give a clear exposition and to facilitate access to the 
individual topics, the sections are rendered 
as self-contained as possible. In section~\ref{sec2}, we review
some of the standard concepts used in  quaternionic quantum mechanics,
i.e. state vector, probability interpretation, scalar product and
left/right quaternionic 
operators~\cite{DEL99a,DIX94,DEL94,DEL96g,ROT89a,REM78}.
Section~\ref{sec3} contains a brief discussion of the momentum 
operator. In section~\ref{sec4}, we summarize without proofs the relevant 
material on quaternionic eigenvalue equations from~\cite{DEL00}. 
Section~\ref{sec5} is devoted
to the study of the one-dimensional Schr\"odinger equation in quaternionic 
quantum mechanics. Sections~\ref{sec6} and \ref{sec7} provide a detailed 
exposition of quaternionic and complex linear differential equations.
In section~\ref{sec8}, we apply the results of previous sections to the
one-dimensional Schr\"odinger equation with quaternionic constant potentials.
Our conclusions are drawn in the final section.


\section{STATES AND OPERATORS IN QUATERNIONIC QUANTUM MECHANICS}
\label{sec2}

In this section, we give a brief survey of the basic mathematical 
tools used in quaternionic quantum 
mechanics~\cite{HOR84,ADL94d,DEL97c,DIX94,ADL95a,GUR96}.  
The quantum state of a particle is
defined, at a given instant,  by a quaternionic wave function  
interpreted as a probability amplitude given by 
\begin{equation}
\Psi (\boldsymbol{r}) = \left[ \, 
f_{\0} + \boldsymbol{h} \cdot \boldsymbol{f} \, \right] (\boldsymbol{r})~,
\end{equation}
where $\boldsymbol{h}=(i,j,k)$,  $\boldsymbol{f}=(f_{\1}, f_{\2},f_{\3})$ and 
$f_{\m}: \mathbb{R}^{\3} \rightarrow \mathbb{R}$, 
{\footnotesize $m=0,1,2,3$}.  
The probabilistic interpretation
of this wave function requires that it belong to 
the Hilbert vector space of square-integrable functions. We shall denote by 
$\mathcal{F}$ the set of wave functions composed of sufficiently regular
functions of this vector space. The same function $\Psi (\boldsymbol{r})$ can be
represented by several distinct sets of components, each one corresponding
to the choice of a particular basis. With each pair of elements of 
$\mathcal{F}$, $\Psi(\boldsymbol{r})$ and $\Phi(\boldsymbol{r})$, we associate the 
quaternionic scalar product 
\begin{equation}
\left( \Psi , \Phi \right) = \int  d^{\3}r \, \, 
\overline{\Psi}(\boldsymbol{r}) \, \Phi(\boldsymbol{r})~,
\end{equation}
where 
\begin{equation} 
\overline{\Psi}(\boldsymbol{r}) = \left[ \, f_{\0} - \boldsymbol{h} \cdot 
\boldsymbol{f} \, \right]
(\boldsymbol{r}) 
\end{equation}
represents the quaternionic conjugate of $\Psi(\boldsymbol{r})$.

A quaternionic linear operator, $\mathcal{O}_{\Ha}$, associates with every
wave function $\Psi(\boldsymbol{r}) \in \mathcal{F}$  another wave function 
$\mathcal{O}_{\Ha} \Psi(\boldsymbol{r}) \in \mathcal{F} $, 
the correspondence being linear from the
right on $\mathbb{H}$
\begin{eqnarray*}
{\cal O}_{\Ha} \, \left[ \, \Psi_{\1}(\boldsymbol{r}) \, q_{\1} + 
\Psi_{\2}(\boldsymbol{r}) \, q_{\2} \,  
\right] & = & 
\left[ \, {\cal O}_{\Ha}  \, \Psi_{\1}(\boldsymbol{r})  \, \right] \, q_{\1} +
\left[ \, {\cal O}_{\Ha}  \, \Psi_{\2}(\boldsymbol{r})  \, \right] \, q_{\2}~,
\end{eqnarray*}
$q_{\1 , \2} \in \mathbb{H}$. Due to the non-commutative nature of the 
quaternionic field we need to  introduce complex and real linear
quaternionic operators,  respectively denoted by 
${\cal O}_{\Co}$ and ${\cal O}_{\Re}$, 
the correspondence being linear  from the right on $\mathbb{C}$
and $\mathbb{R}$
\begin{eqnarray*}
{\cal O}_{\Co} \, \left[ \, \Psi_{\1}(\boldsymbol{r}) \, z_{\1} + 
\Psi_{\2}(\boldsymbol{r}) \, z_{\2} \, 
\right] & = & 
\left[ \, {\cal O}_{\Co}  \, \Psi_{\1}(\boldsymbol{r})  \, \right] \, z_{\1} +
\left[ \, {\cal O}_{\Co}  \, \Psi_{\2}(\boldsymbol{r})  \, \right] \, 
z_{\2}~,\\
{\cal O}_{\Re} \, \left[ \, \Psi_{\1}(\boldsymbol{r}) \, \lambda_{\1} + 
\Psi_{\2}(\boldsymbol{r}) \, \lambda_{\2} 
\, \right] & = &
\left[ \, {\cal O}_{\Ha}  \, \Psi_{\1}(\boldsymbol{r})  \, \right] 
\, \lambda_{\1} +
\left[ \, {\cal O}_{\Ha}  \, \Psi_{\2}(\boldsymbol{r}) \,  \right] 
\, \lambda_{\2}~,
\end{eqnarray*}
$z_{\1 , \2} \in \mathbb{C}$ and $\lambda_{\1 , \2} \in \mathbb{R}$. 

As a concrete illustration of these operators let us consider the case of a
finite, say $n$-dimensional, quaternionic Hilbert space. The wave function
$\Psi(\boldsymbol{r})$ will then be a column vector
\[ 
\Psi = \left( \begin{array}{c} 
\Psi_{\1}\\
\Psi_{\2}\\
\vdots\\
\Psi_{\n} \end{array} \right)~,~~~~~
\Psi_{\1,\2,\dots,\n} \in \mathcal{F}~.
\]
Quaternionic, complex and real linear operators will be
represented by $\nxn$ qua\-ternionic matrices 
$M_{\n}\left[ \boldsymbol{\mathcal{A}} \otimes \mathcal{O} \right]$, where
$\mathcal{O}$ represents the space of real operators acting on the  
components of $\Psi$ and 
$\boldsymbol{\mathcal{A}} = \left( \mathcal{A}_{\Ha} , 
\mathcal{A}_{\Co} , \mathcal{A}_{\Re} \right)$ denote
the real algebras
\begin{eqnarray*} 
\mathcal{A}_{\Ha}~:~
\boldsymbol{\{} \, \boldsymbol{1}~,~~\boldsymbol{L}~,~~\boldsymbol{R}~,
~~\boldsymbol{L} \boldsymbol{*} \boldsymbol{R} \, 
 \boldsymbol{\}}_{\1 \6} ~,\\
\mathcal{A}_{\Co}~:
~\boldsymbol{\{} 
\, \boldsymbol{1}~,~~\boldsymbol{L}~,~~R_{\i}~,~~\boldsymbol{L} \,  
R_{\i} \, \boldsymbol{\}}_{\8}~,\\
\mathcal{A}_{\Re}~:
~\boldsymbol{\{} \, \boldsymbol{1}~,~~\boldsymbol{L}  \, 
\boldsymbol{\}}_{\4}~,
\end{eqnarray*}
generated by the left and right operators  
\begin{equation}
\label{LR}
\boldsymbol{L} := \left(  L_{\i} , L_{\j} , L_{\k}  \right)~,~~~
\boldsymbol{R} := \left(  R_{\i} , R_{\j} , R_{\k}  \right)
\end{equation}
and by the mixed  operators
\begin{equation}
\boldsymbol{L}  \boldsymbol{*} \boldsymbol{R} :=
 \left\{ L_{\p} R_{\q} \right\}~~~~~
\mbox{\footnotesize $p,q=i,j,k$}~.
\end{equation} 
The  action of these operators on the quaternionic wave function 
$\Psi$ is given by 
\[ \boldsymbol{L} \Psi \equiv \boldsymbol{h} \Psi~,~~~ 
\boldsymbol{R} \Psi  \equiv  \Psi \boldsymbol{h}~.
\] 
The operators $\boldsymbol{L}$ and $\boldsymbol{R}$ satisfy the 
left/right quaternionic algebra 
\begin{equation*}
L_{\i}^{\2} = L_{\j}^{\2} =L_{\k}^{\2} =L_{\i} L_{\j} L_{\k} =
R_{\i}^{\2} = R_{\j}^{\2} =R_{\k}^{\2} =R_{\k} R_{\j} R_{\i} =
- \boldsymbol{1}~,
 \end{equation*}
and the following commutation relations
\begin{equation*}
\left[ \, L_{p} \, , \, R_{\q} \, \right] = 0~.
\end{equation*}


\section{SPACE TRANSLATIONS AND QUATERNIONIC MOMENTUM OPERATOR}
\label{sec3}

Space translation operators in quaternionic quantum mechanics are defined 
in the coordinate representation by the real linear anti-hermitian 
operator~\cite{ADL95a} 
\begin{equation}
\boldsymbol{\partial} \equiv \left( \partial_{\x} ,   \partial_{\y}  ,  
\partial_{\z}  \right)~.
\end{equation} 
To construct an observable momentum 
operator we must look for an hermitian operator
that has all the properties of the momentum expected by analogy with the 
momentum operator in complex quantum mechanics. The choice of the 
quaternionic linear operator
\begin{equation}   
\label{pl}
\boldsymbol{\mathcal{P}}_{\L} = - L_{\i} \, \hbar \, \boldsymbol{\partial}~,
\end{equation}
as hermitian momentum operator, would 
appear completely satisfactory, until we consider the translation invariance
for quaternionic Hamiltonians, $\mathcal{H}_{\q}$. 
In fact, due to the presence of 
the  left acting  imaginary unit $i$, the momentum operator~(\ref{pl}) 
does not commute with the $j/k$-part of $\mathcal{H}_{\q}$. 
Thus, although this definition of the 
momentum operator gives an hermitian operator, we must return to the 
anti-hermitian operator $\boldsymbol{\partial}$ to get a translation 
generator,   
$\left[ \boldsymbol{\partial} , \mathcal{H}_{\q} \right] =0$.  
A second possibility to be 
considered is represented by the  complex linear momentum 
operator, introduced by Rotelli in~\cite{ROT89a},   
\begin{equation}
\boldsymbol{\mathcal{P}}_{\R} = - R_{\i} \, \hbar \, \boldsymbol{\partial}~.
\end{equation}
The commutator of $\boldsymbol{\mathcal{P}}_{\R}$ with a quaternionic 
linear operator $\mathcal{O}_{\Ha}$ gives
\[
\left[ \boldsymbol{\mathcal{P}}_{\R} , \mathcal{O} \right] \Psi = \hbar \, 
\left[ \mathcal{O}, \boldsymbol{\partial} \right] \Psi i~.
\]
Taking $\mathcal{O}_{\Ha}$ to be a translation
invariant quaternionic Hamiltonian $\mathcal{H}_{\q}$, we have  
\[ \left[ \boldsymbol{\mathcal{P}}_{\R} , \mathcal{H}_{\q} \right] =0~.
\]
However, this second definition of the momentum operator has the following  
problem: the complex linear momentum operator 
$\boldsymbol{\mathcal{P}}_{\R}$ does not represent a quaternionic hermitian 
operator. In fact, by computing the difference
\[
\left( \Psi , \boldsymbol{\mathcal{P}}_{\R} \Phi \right) - \overline{
\left( \Phi, \boldsymbol{\mathcal{P}}_{\R} \Psi \right)}~, \]
which should vanish for an hermitian operator 
$\boldsymbol{\mathcal{P}}_{\R}$, we find
\begin{equation}
\label{pr}
\left( \Psi , \boldsymbol{\mathcal{P}}_{\R} \Phi \right) -
\left( \boldsymbol{\mathcal{P}}_{\R} \Psi , \Phi \right) = \hbar \,
\left[ i , \left( \Psi , \boldsymbol{\partial} \Phi \right) \right]~,
\end{equation}
which is in general non-vanishing.
There is one important case in which the right-hand side of 
equation~(\ref{pr})
does vanish. The operator $\boldsymbol{\mathcal{P}}_{\R}$  gives a 
satisfactory definition of the  hermitian momentum operator when 
restricted to a 
{\em complex geometry}~\cite{REM78}, that is a {\em complex  projection} of the 
quaternionic scalar product, 
$\left( \Psi , \boldsymbol{\mathcal{P}}_{\R} \Phi \right)_{\Co}$.  
Note that the assumption of a complex projection of the 
quaternionic scalar product  does not imply 
complex wave functions. The state of quaternionic quantum mechanics with
complex geometry will be again described by vectors of a quaternionic Hilbert 
space. In quaternionic quantum mechanics with complex geometry  
observables can be represented by the quaternionic  
hermitian operator, $H$, obtained taking 
the {\em spectral decomposition}~\cite{DEL00} 
of the corresponding anti-hermitian operator, $A$, or simply by the 
complex linear operator, 
$-A R_{\i}$, obtained by multiplying $A$ by
the operator representing the right action of the imaginary unit $i$.  
These two possibilities represent 
equivalent choices in describing 
quaternionic observables within a quaternionic
formulation of quantum mechanics based on complex geometry. In this scenario,
the complex linear 
operator $\boldsymbol{\mathcal{P}}_{\R}$  has all the expected 
properties of the momentum operator. It satisfies the standard commutation 
relations with the coordinates. It is a translation generator.  
Finally, it represents a {\em quaternionic observable}. A review of 
quaternionic and complexified quaternionic quantum mechanics by adopting
a complex geometry is found in~\cite{DEL97c}.


\section{OBSERVABLES IN QUATER\-NIO\-NIC QUAN\-TUM MECHANICS}
\label{sec4}

In a recent paper~\cite{DEL00}, we find a detailed discussion of 
eigenvalue equations within a quaternionic formulation of quantum mechanics
with quaternionic and complex geometry.  Quaternionic eigenvalue equations 
for quaternionic and complex linear 
operators require eigenvalues from the right. In particular, 
without loss of generality, we can 
reduce the eigenvalue problem for quaternionic and complex 
linear anti-hermitian operators $A 
\in M_{\n}\left[ \mathcal{A}_{\Ha} \otimes \mathcal{O} \right]$
to
\begin{equation}
A \, \Psi_{\m} = \Psi_{\m} \, \lambda_{\m} i~~~~~
\mbox{\footnotesize $m=1,2,...,n$}~,
\end{equation}
where $\lambda_{\m}$ are real eigenvalues.   

There is an important difference 
between the structure of hermitian operators in complex and  
quaternionic quantum mechanics. In complex quantum mechanics  
we can always trivially relate an anti-hermitian
operator, $A$, to an hermitian operator, $H$, by 
removing a factor $i$, i.e. $A = i \, H$. In general, 
due to the non-commutative nature of the quaternionic field, this does not 
apply to quaternionic  quantum mechanics.

Let $\left\{ \Psi_{\m} \right\}$ be a set of 
normalized eigenvectors of $A$ with complex imaginary 
eigenvalues $\left\{ i \lambda_{\m} \right\}$. The anti-hermitian operator 
$A$ is then represented by  
\begin{equation} 
\label{oa}
A = \sum_{\r \= \1}^{\n} \Psi_{\r} \, \lambda_{\r}  i \, 
\Psi_{\r}^{\, \sdag}~,
\end{equation}
where $\Psi^{\, \sdag} : = \overline{\Psi}^{\, \t}$. 
It is easy to verify that 
\[ 
A \, \Psi_{\m} = \sum_{\r \= \1}^{\n} \Psi_{\r} \,  \lambda_{\r}  i \, 
\Psi_{\r}^{\, \sdag} \, \Psi_{\m} = 
\sum_{\r \= \1}^{\n} \Psi_{\r} \,  \lambda_{\r}  i \, 
\delta_{\r \m} =
 \Psi_{\m} \,  \lambda_{\m}   i~.
\]
In quaternionic quantum mechanics with quaternionic geometry~\cite{ADL95a},
the observable corresponding to the anti-hermitian operator $A$  
is represented by  
the following hermitian quaternionic linear operator  
\begin{equation}
\label{h} 
H = \sum_{\r \= \1}^{\n} \Psi_{\r}  \,  \lambda_{\r}  \, 
\Psi_{\r}^{\, \sdag}~.
\end{equation}
The action of this operators on the eigenvectors $\Psi_{\m}$ gives 
\[
H \, \Psi_{\m} = \Psi_{\m} \, \lambda_{\m}~.
\]
The eigenvalues of the operator $H$ are real and eigenvectors 
corresponding to different eigenvalues are orthogonal.  

How to relate the hermitian operator $H$ to the anti-hermitian
operator $A$? A simple calculation shows that the operators
$L_{\i} H$ and $ H L_{\i}$ does not satisfy the same
eigenvalue equation of $A$. In fact,
\begin{eqnarray*}
L_{\i} \, H \, \Psi_{\m} & = & 
\left[ L_{\i}  \left( \sum_{\r \= \1}^{\n} \Psi_{\r} \,  \lambda_{\r} \, 
\Psi_{\r}^{\, \sdag} \right) \right]  \Psi_{\m} =   
 i \sum_{\r \= \1}^{\n} \Psi_{\r} \,  \lambda_{\r}  \, 
\Psi_{\r}^{\, \sdag} \, \Psi_{\m} = 
i  \, \Psi_{\m} \, \lambda_{\m} 
\end{eqnarray*}
and
\begin{eqnarray*}
H \, L_{\i} \, \Psi_{\m} & =  &
\left[ \left( \sum_{\r \= \1}^{\n} \Psi_{\r} \,  \lambda_{\r}   \, 
\Psi_{\r}^{\, \sdag} \right) L_{\i} \right]  \Psi_{\m}
 = \sum_{\r \= \1}^{\n} \Psi_{\r} \,  \lambda_{\r}  \, 
\Psi_{\r}^{\, \sdag} i \, \Psi_{\m}~.
\end{eqnarray*}
These  problems can be avoided by using the right operator $R_{\i}$ 
instead of the left operator $L_{\i}$. In fact, the operator  
$H R_{\i}$ satisfies the same eigenvalue equation of $A$, 
\begin{eqnarray*}
H \, R_{\i} \Psi_{\m} &=& 
\left[ \left( \sum_{\r \= \1}^{\n} \Psi_{\r} \,  \lambda_{\r}   \, 
\Psi_{\r}^{\, \sdag} \right) R_{\i} \right] \Psi_{\m} =  
\sum_{\r \= \1}^{\n} \Psi_{\r} \,  \lambda_{\r}   \, 
\Psi_{\r}^{\, \sdag} \Psi_{\m} i = 
\Psi_{\m} \, \lambda_{\m} i~.
\end{eqnarray*}
The eigenvalues of the operator $-A R_{\i}$ are real and 
eigenvectors corresponding to different eigenvalues are orthogonal. 
The right hermiticity of this operator is recovered  within a  
quaternionic formulation of quantum mechanics based 
on complex geometry~\cite{DEL97c}.

When the space state is finite-dimensional, it is always possible to form a 
basis with the eigenvectors of the operators $H$ and  $-A R_{\i}$.  
When the space state is infinite-dimensional, this is no longer necessarily 
the case. So, it is useful to introduce a new concept, that of an 
observable. By definition, the hermitian operators $H$ or 
$-A R_{\i}$ are observables if the orthonormal system of 
vectors forms a basis in the state space. 

In quaternionic quantum 
mechanics with quaternionic geometry, the hermitian operator corresponding 
to the anti-hermitian operator $A$ of equation (\ref{oa}) is thus given by 
the operator $H$  of equation (\ref{h}). 
By adopting a  complex geometry, observables can also be 
represented by complex linear  hermitian operators  obtained by multiplying
the corresponding anti-hermitian operator $A$  by $-R_{\i}$. 
We remark that for complex eigenvectors, the operators 
 $L_{\i} H$, $H L_{\i}$, $H R_{\i}$ and $A$ reduce to the same complex 
operator
\begin{equation*} 
i \, H = i \, \sum_{\r \= \1}^{\n} \lambda_{\r}  \,  \Psi_{\r} \, 
\Psi_{\r}^{\, \sdag} ~.
\end{equation*}

We conclude this section by giving an explicit example of 
quaternionic hermitian operators 
in a finite two-dimensional space state. Let
\begin{equation}
\label{s1}
A = 
\left( \begin{array}{cc} $-$ i & ~3 j \\ 3 j & ~i \end{array} \right)
\end{equation}
be an anti-hermitian operator. An easy computation shows that the 
eigenvalues and the eigenvectors of this operator are given by
\[   
\left\{ 2 i \, , \, 4 i \right\}~~~
\mbox{and}~~~ 
\left\{ \, \mbox{$\frac{1}{\sqrt{2}}$} \, 
\left( \begin{array}{c} i \\ j \end{array} 
\right) \, , \, \mbox{$\frac{1}{\sqrt{2}}$} \, 
~\left( \begin{array}{c} k \\ 1 \end{array} \right) \, \right\}~.
\]
It is immediate to verify that $iA$ and $Ai$ are characterized by complex 
eigenvalues and so cannot represent quaternionic
observables. In quaternionic quantum mechanics with quaternionic geometry, 
the quaternionic observable corresponding to the anti-hermitian 
operator of equation (\ref{s1}) is  given by 
the hermitian operator 
\begin{equation}
H = \Psi_{\1} \, 2 \, \Psi_{\1}^{\sdag} +  
\Psi_{\2} \, 4 \, \Psi_{\2}^{\sdag} =  
\left( \begin{array}{cc} 3 & ~k \\ $-$ k & ~3 \end{array} \right)~.
\end{equation}
Within a quaternionic quantum mechanics with complex geometry, a second
equivalent definition  of  the quaternionic observable corresponding to 
the anti-hermitian  operator of equation (\ref{s1})
is given  by the complex linear hermitian  operator 
\begin{equation}
\tilde{H} = 
\left( \begin{array}{cc} $-$ i & ~3 j \\ 3 j & ~i \end{array} \right)
\, R_{\i}~.
\end{equation}


\section{THE QUATERNIONIC SCHR\"ODINGER EQUATION}
\label{sec5}

For simplicity, we shall assume a one-dimensional description. 
In the standard formulation of quantum mechanics, the wave function of a
particle whose potential energy  is $V(x,t)$ must satisfy 
the Schr\"odinger equation 
\begin{eqnarray}
\label{cse}
i \, \hbar \, \partial_{\t} \Phi(x,t) & = & \mathcal{H}
 \, \Phi(x,t) 
=   \left[ - \mbox{$\frac{\hbar^{\2}}{2m}$} \, \partial_{\xx} 
+ V(x,t) \right] \Phi(x,t)~.
\end{eqnarray}
Let us modify the previous equation by introducing the quaternionic potential
\[
\left[ V +  \boldsymbol{h} \cdot  \boldsymbol{V} \right] (x,t)~.
\] 
The $i$-part of this quaternionic potential violates
the norm conservation. In fact,   
\[
\begin{array}{lcl}
\partial_{\t} \, 
\int_{\infm}^{\infp} 
 \mbox{d} x \, \overline{\Phi}  \, \Phi   & = & 
\int_{\infm}^{\infp} 
\mbox{d} x \, \left[ 
\mbox{$\frac{\hbar}{2m}$} \, 
\overline{\Phi}  \, i \, \partial_{\xx} \Phi - 
\mbox{$\frac{\hbar}{2m}$} \, \left( \partial_{\xx} \overline{\Phi} \right) 
\, i \, \Phi
 - \mbox{$\frac{1}{\hbar}$} \, \overline{\Phi}  
\left\{ i ,  \boldsymbol{h} \right\}  \cdot  \boldsymbol{V} \Phi
\right] 
 = 
\mbox{$\frac{2}{\hbar}$} \, 
\int_{\infm}^{\infp} 
\mbox{d} x \,
 \overline{\Phi} \, V_{\1} \Phi~.
\end{array}
\]
The $j/k$-part of $\boldsymbol{h} \cdot  \boldsymbol{V}$  is 
 responsible for T-violation~\cite{ADL88}.  To show that, 
we briefly discuss  the  
time reversal invariance in quaternionic quantum mechanics. 
The quaternionic Schr\"odinger equation in  
presence of a quaternionic potential
which preserves norm conservation, is given by~\cite{ADL88,DAV89,DAV92,ADL95a}
\begin{equation}
\label{qse}
i \, \hbar \, \partial_{\t} \Phi(x,t) = 
\left[ \,  \mathcal{H} - j \, W \, \right] \, \Phi(x,t)~,
\end{equation}
where $W \in \mathbb{C}$. Evidently, quaternionic conjugation 
\begin{equation*}
- \, \hbar \, \partial_{\t} \overline{\Phi}(x,t) \, i = \mathcal{H} \, 
\overline{\Phi}(x,t)
+ \overline{\Phi}(x,t) \,  j \, W 
\end{equation*}
does not yield a time-reversed version of the original Schr\"odinger 
equation 
\begin{equation}
\label{te}
- \, i \, \hbar \, \partial_{\t} \Phi_{\T}(x,-t)  = 
\left[ \, \mathcal{H} - j \, W \,  \right] \, \Phi_{\T}(x,-t)~.
\end{equation}
To understand why the T-violation is proportional to the $j/k$-part of the
quaternionic potential,  let us 
consider a real potential $W$.   
Then, the  Schr\"odinger equation has a T-invariance. 
By multiplying the equation (\ref{qse}) by $j$ from the left, we have
\begin{equation*}
- \, i \, \hbar \, \partial_{\t} \, j \, \Phi(x,t) =  
\left[ \, \mathcal{H} - j \, W \, \right] \, j \, \Phi(x,t)~,~~~
W \in \mathbb{R}~,
\end{equation*}
which has the same form of equation (\ref{te}). Thus,
\[ \Phi_{\T}(x,-t) = j \, \Phi(x,t)~.\]
A similar discussion applies for imaginary complex potential 
$W \in i \, \mathbb{R}$. 
In this case, we find
\[ \Phi_{\T}(x,-t) = k \, \Phi(x,t)~.\]
However, when both $V_{\2}$ and $V_{\3}$ are non zero, i.e 
$W \in \mathbb{C}$,  this construction 
does not work, and the quaternionic physics is T-violating. The system
of neutral kaons is the natural candidate to study the presence 
of {\em effective} quaternionic potentials, 
$V+ \boldsymbol{h} \cdot  \boldsymbol{V}$. In studying such a system, we need 
of $V_{\1}$ and  $V_{\2 , \3}$ in order to include the decay rates of 
$K_{\Sh}$/$K_{\Lo}$ and CP-violation effects.

\subsection{Quaternionic stationary states}

For stationary states, 
\[V(x,t)=V(x)~~~\mbox{and}~~~W(x,t)=W(x)~,\]
we look for
solutions of the Schr\"odinger equation of the form
\begin{equation}
\label{sepvar}
\Phi(x,t) = \Psi(x) \, \zeta(t)~.
\end{equation}
Substituting (\ref{sepvar}) in the quaternionic Schr\"odinger equation,
we obtain 
\begin{equation}
i \, \hbar \, \Psi(x) \, \dot{\zeta}(t) = \left[ \, \mathcal{H} 
- j \, W(x) \, \right]
\, \Psi(x) \, \zeta(t)~.
\end{equation}
Multiplying by $- \overline{\Psi}(x) \, i$ from the left and by 
$\overline{\zeta}(t)$ from the right, we find
\begin{equation}
 \label{es1}
\hbar \, \dot{\zeta}(t) \, \overline{\zeta}(t)
\, / \, | \zeta(t) |^{\2} = \overline{\Psi}(x) \,   
\left[ \,  -  \, i \, \mathcal{H}  + k \, W(x) \, \right] \, \Psi(x) 
\, / \, | \Psi(x) |^{\2} ~.
\end{equation}
In this equation we have a function of $t$ in the left-hand side and a 
function of $x$ in the  right-hand side. The previous equality is only 
possible if 
\begin{equation}
\label{ess}
 \hbar \, \dot{\zeta}(t) \, \overline{\zeta}(t)
\, / \, | \zeta(t) |^{\2} = \overline{\Psi}(x) \,   
\left[ \,  -  \, i \, \mathcal{H}  + k \, W(x) \, \right] \, \Psi(x) 
\, / \, | \Psi(x) |^{\2} = q ~,
\end{equation}
where $q$ is a quaternionic constant. The energy operator 
$- i \, \mathcal{H}  + k \, W(x)$ represents an 
anti-hermitian operator. Consequently, its eigenvalues are purely 
imaginary  quaternions, $q=\boldsymbol{h} \cdot \boldsymbol{E}$. 
By applying the unitary 
transformation $u$, 
\[ \overline{u}  \, \, \boldsymbol{h} \cdot \boldsymbol{E} \, \, u = - \, i \, 
E~,~~~~~
E = \sqrt{E_{\1}^{\2} +   E_{\2}^{\2} +   E_{\3}^{\2}}~,
\]
equation (\ref{ess}) becomes
\begin{equation}
\label{es2}
\hbar \, \overline{u} \, \dot{\zeta}(t) \, \overline{\zeta}(t) \, u 
\, / \, | \zeta(t) |^{\2} =  
\overline{u} \, \overline{\Psi}(x) \,  
\left[\,  - \, i \, H  + k \, W(x) \, \right] \, \Psi(x) \, u \,  
/ \, | \Psi(x) |^{\2} 
= - \, i \, E~.
\end{equation}
The solution $\Phi(x,t)$ of the Schr\"odinger equation is not modified
by this similarity transformation. In fact, 
\[ \Phi(x,t) \rightarrow \Psi(x) \, u \, \overline{u} \, \zeta(t) =
\Psi(x) \, \zeta(t)~.
\] 
By observing that  $|\Phi(x,t)|^{\2} = |\Psi(x)|^{\2} 
|\zeta(t)|^{\2}$, the norm conservation implies  
$|\zeta(t)|^{\2}$ constant. Without loss of generality, we can choose
$|\zeta(t)|^{\2}=1$. Consequently, by equating the first and the third term  
in equation (\ref{es2}) and solving the corresponding equation, 
we  find
\begin{equation}
\label{chi}
\zeta(t) =  \exp [-i Et / \hbar] \, \zeta(0)~,
\end{equation}
with $\zeta(0)$ unitary quaternion. Note that the position of $\zeta(0)$
in equation (\ref{chi}) is very important. In fact, it can be shown that  
$\zeta(0)\, \exp [-i Et / \hbar]$  is  not solution of
equation (\ref{es2}). Finally, to complete the solution of 
the quaternionic Schr\"odinger equation,
we must determine $\Psi(x)$ by solving the following second order 
(right complex linear) differential equation
\begin{equation}
\label{psi} 
\left[ \, i \, \mbox{$\frac{\hbar^{\2}}{2m}$} \, \partial_{\xx} 
- i \,  V(x)  + k \, W(x) \, \right] \Psi(x)   
= - \Psi(x)  \, i \, E~.
\end{equation}

\subsection*{$\bullet$ Real potential}

For $W(x)=0$, equation (\ref{psi}) becomes
\begin{equation}
\label{psi2} 
\left[ \, \mbox{$\frac{\hbar^{\2}}{2m}$} \, \partial_{\xx} 
- V(x) \, \right] \, \left\{ [\Psi(x)]_{\Co} - j \,   [j \Psi(x)]_{\Co} 
\right\}  
= i \, 
 \left\{ [\Psi(x)]_{\Co} - j \,   [j \Psi(x)]_{\Co} 
\right\}  
\, i \, E~.
\end{equation}
Consequently,
\[
\left[ \, \mbox{$\frac{\hbar^{\2}}{2m}$} \, \partial_{\xx} 
-   V(x)  \, \right] \, [\Psi(x)]_{\Co}   
= - [\Psi(x)]_{\Co} \, E~,
\]
and
\[
\left[ \, \mbox{$\frac{\hbar^{\2}}{2m}$} \, \partial_{\xx} 
-   V(x)  \right] \, [j \Psi(x)]_{\Co}   
= [j \Psi(x)]_{\Co} \, E~.
\]
By solving these complex equations, we find 
\begin{eqnarray*}
\Psi(x) & = & \exp \left[ \mbox{\footnotesize 
$\sqrt{ \mbox{$\frac{2m}{\hbar^{\2}}$} \, 
\left( V - E \right)}$} \, x \right] \, k_{\1} +
  \exp \left[ -  \mbox{\footnotesize 
$\sqrt{ \mbox{$\frac{2m}{\hbar^{\2}}$} \, 
\left( V - E \right)}$} \, x \right] \, k_{\2} + \\
 & & j \, \left\{ \, 
 \exp \left[ \mbox{\footnotesize 
$
\sqrt{ \mbox{$\frac{2m}{\hbar^{\2}}$} \, 
\left( V + E \right)}$} \, x \right] \, k_{\3} +
  \exp \left[ -  
\mbox{\footnotesize 
$\sqrt{ \mbox{$\frac{2m}{\hbar^{\2}}$} \, 
\left( V + E \right)}$} \, x \right] \, k_{\4} \, 
\right\}~,
\end{eqnarray*}
where $k_{\n}$, {\footnotesize $n=1,...,4$}, 
 are complex coefficients determined by the initial 
conditions.

\subsection*{$\bullet$ Free particles}

For free particles, $V(x)=W(x)=0$, the previous solution reduces 
to
\begin{eqnarray*}
\Psi(x) & = & \exp \left[ \, i \,  
\mbox{$\frac{p}{\hbar}$} \, x \, \right] \, k_{\1} +
exp \left[ \, -  i \,  
\mbox{$\frac{p}{\hbar}$} \, x \, \right] \, k_{\2} + j \, \left\{ 
\exp \left[ \,  
\mbox{$\frac{p}{\hbar}$} \, x \, \right]
\, k_{\3} +
 \exp \left[ \, -   
\mbox{$\frac{p}{\hbar}$} \, x \,  \right] \, k_{\4}
\right\}~,
\end{eqnarray*}
where $p = \sqrt{2mE}$. 
For scattering problems with a wave function incident from the left 
on quaternionic potentials, we have 
\begin{equation}
\label{psii}
\Psi(x) = 
\exp [ \,  i \,  
\mbox{$\frac{p}{\hbar}$} \, x \, ] + r \, \exp[ \, - i \,  
\mbox{$\frac{p}{\hbar}$} \, x \, ] + 
j \, \tilde{r} \, \exp[ \,  
\mbox{$\frac{p}{\hbar}$} \, x \,]~,
\end{equation}
where $|r|^{\2}$ is the standard coefficient of reflection and 
$|\tilde{r} \, \exp[  \,  
\mbox{$\frac{p}{\hbar}$} \, x \, ]|^{\2}$ represents an additional  evanescent 
probability of reflection. In our study of quaternionic potentials, we 
shall deal with
the rectangular potential  barrier of width $a$. In this case, the particle 
is free for $x<0$, where the solution is given by (\ref{psii}), and $x>a$,
where the solution is
\begin{equation}
\label{psit}
\Psi(x) = 
t \, \exp[ \, i \,  
\mbox{$\frac{p}{\hbar}$} \, x \, ] + 
j \, \tilde{t} \, \exp[ \, - \,  
\mbox{$\frac{p}{\hbar}$} \, x \,]~.
\end{equation}
Note that, in equations (\ref{psii}) and 
(\ref{psit}), we have respectively omitted the 
complex exponential solution $\exp[  \, - 
\mbox{$\frac{p}{\hbar}$} \, x \, ]$ and 
$\exp[  \, 
\mbox{$\frac{p}{\hbar}$} \, x \, ]$
which are
in conflict with the boundary condition that $\Psi(x)$ remain finite as
$x \to - \infty$ and $x \to \infty$. In equation (\ref{psit}),  we have also 
omitted the complex exponential solution $\exp[ \, - i \,  
\mbox{$\frac{p}{\hbar}$} \, x \, ]$ because we are considering a wave incident
from the left.


\section{QUATERNIONIC LINEAR DIFFERENTIAL EQUATION}
\label{sec6}

Consider the second order quaternionic linear differential operator
\[
\mathcal{D}_{\Ha} = \partial_{\xx} + \left( 
a_{\0} + \boldsymbol{L} \cdot \boldsymbol{a} \right) \partial_{\x}
+  b_{\0} + \boldsymbol{L} \cdot \boldsymbol{b}~~~\in
\mathcal{A_{\Ha}} \otimes \mathcal{O}~.
\]
We are interested in finding the solution of the 
quaternionic linear differential 
equation
\begin{equation}
\label{qlde}
\mathcal{D}_{\Ha} \, \varphi(x) = 0~.
\end{equation} 
In analogy to the complex case, we look for solutions of exponential form
\[ \varphi(x) = \exp[ q x]~, \] 
where  $q \in \mathbb{H}$ and $x \in \mathbb{R}$. 
To satisfy equation  (\ref{qlde}), the 
constant $q$ has to be a solution of the quaternionic quadratic 
equation~\cite{NIV41,NIV42,BRA42,EIL44}   
\begin{equation}
\label{qua}
q^{\2} +  \left( a_{\0} + \boldsymbol{h} \cdot \boldsymbol{a} \right) 
q + b_{\0} + \boldsymbol{h} \cdot \boldsymbol{b} = 0~.
\end{equation}

\subsection{Quaternionic quadratic equation}
\label{s61}

To simplify our discussion,  it is convenient to  modify equation (\ref{qua}) 
by removing the real constant $a_{\0}$.  To do this, we introduce a 
new quaternionic constant  $p$ defined by  
$p = q + \frac{a_{\0}}{2}$. The quadratic equation (\ref{qua}) then 
becomes
\begin{equation}
\label{qua2}
p^{\2} +  \boldsymbol{h} \cdot \boldsymbol{a} \,  
p + c_{\0} + \boldsymbol{h} \cdot \boldsymbol{c} = 0~,
\end{equation}
where $c_{\0} = b_{\0} - \frac{a_{\0}^{\2}}{4}$ and 
$\boldsymbol{c} = \boldsymbol{b} - \frac{a_{\0}}{2} \, \boldsymbol{a}$. 
We shall give the solution of equation (\ref{qua2}) in terms  
of real constant $c_{\0}$ and of the 
real vectors $\boldsymbol{a}$ and $\boldsymbol{c}$. 
Let us  analyze the following cases:\\ 

\begin{tabular}{cl}
$\bullet$   & $\boldsymbol{a} \neq 0$,  
$\boldsymbol{c} \neq 0$:
\end{tabular}
\begin{tabular}{cl}
{\bf (i)}           & 
$\boldsymbol{a} \times   \boldsymbol{c} = 0$,\\
{\bf (ii)} & $\boldsymbol{a} \cdot \boldsymbol{c} = 0$,\\
{\bf (iii)} & 
$\boldsymbol{a} \times   \boldsymbol{c} \neq 0 \neq
\boldsymbol{a} \cdot \boldsymbol{c}$;
\end{tabular}

\begin{tabular}{cl}
$\bullet$  & $\boldsymbol{a} = 0$,  
$\boldsymbol{c} \neq 0$;\\
$\bullet$  & $\boldsymbol{a} \neq 0$,  
$\boldsymbol{c} = 0$;\\
$\bullet$  & $\boldsymbol{a} = \boldsymbol{c} = 0$.
\end{tabular}

\vspace*{0.3cm}

\noindent $\bullet$ {\bf (i)} $\boldsymbol{a} \times   \boldsymbol{c} = 0$. 
In this case   $\boldsymbol{a}$ and $\boldsymbol{c}$ are parallel vectors, so
equation (\ref{qua2}) can be easily reduced
to a complex equation. In fact, by introducing the imaginary unit  
$\mathcal{I} =  \boldsymbol{h} \cdot \boldsymbol{a} / |\boldsymbol{a}|$
and observing that $\boldsymbol{h} \cdot \boldsymbol{c} = \mathcal{I} \, 
\alpha$, with $\alpha \in \mathbb{R}$, we find   
\[
p^{\2} +  \mathcal{I} \, |\boldsymbol{a}| \, 
p + c_{\0} + \mathcal{I} \, \alpha= 0~,
\]
whose {\em complex} solutions are immediately found.

\vspace*{0.3cm}

\noindent $\bullet$ {\bf (ii)} $\boldsymbol{a} \cdot  \boldsymbol{c} = 0$.
By observing that $\boldsymbol{a}$, $\boldsymbol{c}$ and
$\boldsymbol{a} \times \boldsymbol{c}$ are orthogonal vectors, we can 
rearrange the imaginary part of  $p$, 
$\boldsymbol{h} \cdot \boldsymbol{p} $,  in terms of the new basis 
$\left( \boldsymbol{a} , \boldsymbol{c} , \boldsymbol{a} \times 
\boldsymbol{c} \right)$, i.e.    
\begin{equation}
\label{sol2}
p = p_{\0} + \boldsymbol{h} \cdot  
\left( x\, \boldsymbol{a} + y\, \boldsymbol{c} + z\, 
\boldsymbol{a} \times \boldsymbol{c}  \right)~.
\end{equation}
Substituting (\ref{sol2}) in equation (\ref{qua2}), we obtain 
the following system of equations for the real variables  
$p_{\0}$, $x$, $y$ and $z$,\\

\begin{tabular}{ll}
$\mathbb{R}$~:           & 
~~$p_{\0}^{\2} - ( x^{\2} + x ) \, |\boldsymbol{a}|^{\2} -
y^{\2} \, |\boldsymbol{c}|^{\2}  - z^{\2} \, |\boldsymbol{a}|^{\2} 
|\boldsymbol{c}|^{\2} + c_{\0}=0$ 
~,\\
$\boldsymbol{h} \cdot \boldsymbol{a}$~: & 
~~$ p_{\0} \left( 1 + 2 \, x \right)=0$~,\\
$\boldsymbol{h} \cdot \boldsymbol{c}$~: & 
~~$ 1 + 2 \, p_{\0} \, y  - z \, |\boldsymbol{a}|^{\2}=0 $~,\\
$\boldsymbol{h} \cdot \boldsymbol{a} \times \boldsymbol{c}$~: & 
~~$ y + 2 \, p_{\0} \, z=0$~.
\end{tabular}

\vspace*{0.2cm}

\noindent The second equation, $p_{\0} \left( 1 + 2 \, x \right)=0$,
implies $p_{\0}=0$ and/or $x = -\frac{1}{2}$. For $p_{\0}=0$,
it can be shown that the solution of
equation (\ref{qua2}), in terms
of $p_{\0}$, $x$, $y$ and $z$,  is given by 
\begin{equation}
\label{d1}
p_{\0} = 0~,~~~
x = - \mbox{$\frac{1}{2}$}  \pm \sqrt{\Delta}~,~~~
y = 0~,~~~
z = \frac{1}{| \boldsymbol{a} |^{\2}}~, 
\end{equation}
where
\[
\Delta = \frac{1}{4} + \frac{1}{|\boldsymbol{a}|^{\2}} \left( c_{\0} -  
\frac{|\boldsymbol{c}|^{\2}}{|\boldsymbol{a}|^{\2}} \right) \geq 0~.
\]
For $x = -\frac{1}{2}$, we find   
\begin{equation}
\label{d2}
y = - \frac{2 \, p_{\0}}{4 \, p_{\0}^{\2} + | \boldsymbol{a} |^{\2}}
~,~~~
z = \frac{1}{4 \,  p_{\0}^{\2} + | \boldsymbol{a} |^{\2}}~, 
\end{equation} 
and 
\[
p_{\0}^{\2} = \mbox{$\frac{1}{4}$} \, \left[
\pm \, 2 \, \sqrt{c_{\0}^{\2} + | \boldsymbol{c} |^{\2}} - 2 \, c_{\0} 
-  | \boldsymbol{a} |^{\2} \right]~.
\]
It is easily verified that 
\[
\Delta \leq 0 ~\Rightarrow~
\sqrt{c_{\0}^{\2} + | \boldsymbol{c} |^{\2}} - c_{\0} \geq 
\frac{ | \boldsymbol{a} |^{\2}}{2}~,
\]
thus
\begin{equation}
p_{\0} = \pm \, \mbox{$\frac{1}{2}$} \, \sqrt{
2 \left(   \sqrt{c_{\0}^{\2} + | \boldsymbol{c} |^{\2}} - c_{\0} \right)
-  | \boldsymbol{a} |^{\2}}~.
\end{equation}
Summarizing, for $\Delta \neq 0$, we have two quaternionic solutions,
$p_{\1} \neq p_{\2}$, 
\begin{eqnarray}
\label{t1}
\Delta > 0~: &~~~
p_{\0} = 0~, \nonumber\\
 &~~~
x = - \mbox{$\frac{1}{2}$}  \pm \sqrt{\Delta}~, \nonumber \\ 
& ~~~y = 0~, \nonumber \\ 
& ~~~z = \frac{1}{| \boldsymbol{a} |^{\2}}~;\\ 
\label{t11}
\Delta < 0~: &~~~
p_{\0} = \pm \, \mbox{$\frac{1}{2}$} \, \sqrt{
2 \left(   \sqrt{c_{\0}^{\2} + | \boldsymbol{c} |^{\2}} - c_{\0} \right)
-  | \boldsymbol{a} |^{\2}}~, \nonumber \\ 
 & ~~~
x = - \mbox{$\frac{1}{2}$}~, \nonumber \\ 
& ~~~
y = - \frac{2 \, p_{\0}}{4 \, p_{\0}^{\2} + | \boldsymbol{a} |^{\2}}
~, \nonumber \\ 
& ~~~
z = \frac{1}{4 \,  p_{\0}^{\2} + | \boldsymbol{a} |^{\2}}~. 
\end{eqnarray}
For $\Delta = 0$, these solutions tend to the same solution
$p_{\1}=p_{\2}$ given by
\begin{equation}
\label{sc}
\Delta = 0~:~~~
p_{\0} = 0~,~~~
x = - \mbox{$\frac{1}{2}$}~,~~~ 
y = 0~,~~~ z = \frac{1}{| \boldsymbol{a} |^{\2}}~.
\end{equation}

\vspace*{0.3cm}

\noindent $\bullet$ 
{\bf (iii)} 
$\boldsymbol{a} \times   \boldsymbol{c} \neq 0 
\neq \boldsymbol{a} \cdot \boldsymbol{c}$.
In discussing this case, we 
introduce the vector 
$ \boldsymbol{d} = 
\boldsymbol{c} - d_{\0} \, \boldsymbol{a}$, 
$d_{\0} = \boldsymbol{a} \cdot \boldsymbol{c} / 
|\boldsymbol{a}|^{\2}$ and the imaginary part of $p$ 
in terms of the
orthogonal vectors $\boldsymbol{a}$, $\boldsymbol{d}$ and
$\boldsymbol{a} \times \boldsymbol{d}$, 
\begin{equation}
\label{sol3}
p = p_{\0} + \boldsymbol{h} \cdot  
\left( x\, \boldsymbol{a} + y\, \boldsymbol{d} + z\, 
\boldsymbol{a} \times \boldsymbol{d}  \right)~.
\end{equation}
By using this decomposition, from equation (\ref{qua2}) we obtain the following
system of real equations \\

\begin{tabular}{ll}
$\mathbb{R}$~:           & 
~~$p_{\0}^{\2} - ( x^{\2} + x ) \, |\boldsymbol{a}|^{\2} -
y^{\2} \, |\boldsymbol{d}|^{\2}  - z^{\2} \, |\boldsymbol{a}|^{\2} 
|\boldsymbol{d}|^{\2} + c_{\0} =0$ 
~,\\
$\boldsymbol{h} \cdot \boldsymbol{a}$~: & 
~~$ p_{\0} \left( 1 + 2 \, x \right) + d_{\0}=0$~,\\
$\boldsymbol{h} \cdot \boldsymbol{d}$~; & 
~~$ 1 + 2 \, p_{\0} \, y  - z \, |\boldsymbol{a}|^{\2}=0 $~,\\
$\boldsymbol{h} \cdot \boldsymbol{a} \times \boldsymbol{d}$~: & 
~~$ y + 2 \, p_{\0} \, z=0$~.
\end{tabular}

\vspace*{0.2cm}

\noindent The second equation of this system,  
$p_{\0} \left( 1 + 2 \, x \right) + d_{\0}=0$, implies $p_{\0} \neq 0$
since $d_{\0} \neq 0$. Therefore, we have
\begin{equation}
\label{t2}
x = - \frac{p_{\0} + d_{\0}}{2 p_{\0}}~,~~~
y = - \frac{2 p_{\0}}{4 p_{\0}^{\2} + | \boldsymbol{a} |^{\2}}~,~~~
z = \frac{1}{4 p_{\0}^{\2} + | \boldsymbol{a}|^{\2}}~, 
\end{equation}
and 
\begin{equation}
\label{cub}
16 \, w^{\3} + 8 \left[ | \boldsymbol{a} |^{\2} + 2c_{\0} \right] 
w^{\2} + 4 \left[ | \boldsymbol{a} |^{\2} 
(c_{\0} - d_{\0}^{\2}) + \frac{| \boldsymbol{a}|^{\4}}{4} -  
|\boldsymbol{d}|^{\2} \right] 
w - d_{\0}^{\2} 
\, |\boldsymbol{a}|^{\4} = 0~,
\end{equation}
where $w=p_{\0}^{\2}$. By using the Descartes rule of signs it can be proved
that equation (\ref{cub}) has only one real positive solution~\cite{NIV41}, 
$w = \alpha^{\2}$, $\alpha \in \mathbb{R}$. This implies
$p_{0} = \pm \, \alpha$.  Thus, we also find two quaternionic solutions.

\vspace*{.3cm}

\noindent $\bullet$ $\boldsymbol{a} = 0$ and $\boldsymbol{c} \neq 0$.  
By introducing the imaginary {\em complex} unit
$\mathcal{I} =  \boldsymbol{h} \cdot \boldsymbol{c} / |\boldsymbol{c}|$,
we can reduce equation (\ref{qua2}) to the following {\em complex} equation 
\[
p^{\2} + c_{\0} + \mathcal{I} \, |\boldsymbol{c}| = 0~.
\]

\vspace*{.3cm}

\noindent $\bullet$ $\boldsymbol{a} \neq 0$ and $\boldsymbol{c} = 0$.  This
case is similar to the previous one. We introduce the imaginary {\em complex} 
unit $\mathcal{I} =  \boldsymbol{h} \cdot \boldsymbol{a} / |\boldsymbol{a}|$
and reduce equation (\ref{qua2}) to the {\em complex} equation 
\[
p^{\2} + \mathcal{I} \, |\boldsymbol{a}| \, p + c_{\0} = 0~.
\]

\vspace*{.3cm}

\noindent $\bullet$ $\boldsymbol{a} = \boldsymbol{c} = 0$. 
Equation (\ref{qua2}) becomes
\[
p^{\2} + c_{\0} = 0~.
\]
For $c_{\0} = - \alpha^{\2}$, $\alpha \in \mathbb{R}$, 
we find  two real solutions. For $c_{\0} = \alpha^{\2}$, we obtain 
an {\em infinite} number of quaternionic solutions, i.e.  
$p = \boldsymbol{h} \cdot \boldsymbol{p}$, 
where $|\boldsymbol{p}| = |\alpha|$.  

Let us resume our discussion on quaternionic linear quadratic equation.
For $\boldsymbol{a}=0$ and/or $\boldsymbol{c}=0$ and for 
$\boldsymbol{a} \times \boldsymbol{c}=0$ we can reduce
quaternionic linear quadratic equations to  {\em complex} equations.
For non null vectors satisfying $\boldsymbol{a} \cdot  \boldsymbol{c} =0$
or 
$\boldsymbol{a} \times   \boldsymbol{c} \neq 0 
\neq \boldsymbol{a} \cdot \boldsymbol{c}$,
we have {\em effective} quaternionic equations. 
In these cases, we always find two quaternionic solutions 
(\ref{t1}), (\ref{t11}) and (\ref{t2}-\ref{cub}).  
For  $\boldsymbol{a} \cdot \boldsymbol{c} = 0$ 
and $\Delta = 0$,  these solutions tend to the same solution (\ref{sc}). 
Finally, the fundamental theorem of algebra is lost 
 for a {\em restricted} class of quaternionic quadratic linear equations, 
namely
\[ q^{\2} + \alpha^{\2} = 0~,~~~\alpha \in \mathbb{R}~.\]


\subsection{Second order quaternionic differential equations 
with constant coefficients}

Due to the quaternionic linearity from the right of equation (\ref{qlde}), 
we look for general solutions which are of the form 
\[
\varphi(x) = \varphi_{\1} (x) \, c_{\1} + \varphi_{\2} (x) \, c_{\2}~,
\]
where $\varphi_{\1} (x)$ and $\varphi_{\2} (x)$ represent two linear 
independent 
solutions of equation (\ref{qlde}) and
$c_{\1}$ and $c_{\2}$ are quaternionic constants fixed by the initial 
conditions. In analogy to the complex case, we can distinguish between 
quaternionic linear dependent and independent solutions by 
constructing a Wronskian functional. To do this, we need to  define
a  quaternionic determinant. Due to the non-commutative nature of quaternions,
the standard definition of determinant must be revised. 
The study  of quaternionic, complex and real functionals, extending
the complex determinant to quaternionic matrices, has been
extensively developed  in quaternionic linear 
algebra~\cite{BRE51,BRE68,DYS72,CHE91}. 
In a recent paper~\cite{DEL00b}, we find an interesting discussion 
on the impossibility to obtain a quaternionic functional with the main 
properties of the complex determinant. For quaternionic 
matrices, $M$, a {\em real positive} functional, 
$|\mbox{det} M| = \sqrt{\mbox{det} [MM^{\+}]}$, which reduces to the 
absolute value of  the standard determinant for complex matrices, was
introduced by Study~\cite{STU20} and its properties axiomatized by 
Dieudonn\'e~\cite{DIE43}. The details can be found in the excellent survey 
paper of Aslaksen~\cite{ASL96}. 
This functional allows to construct a real positive Wronskian~\cite{DEL00}
\begin{eqnarray*}
\mathcal{W}(x) & = & \left| \mbox{det} \left( \begin{array}{cc}
                \varphi_{\1} (x) & ~\varphi_{\2} (x)\\
                \dot{\varphi}_{\1} (x) & ~\dot{\varphi}_{\2} (x) 
              \end{array}  \right) \right| \\
 & = &
| \varphi_{\1}(x) | \, | \dot{\varphi}_{\2}(x) - \dot{\varphi}_{\1}(x)
\, \varphi_{\1}^{- \1}(x) \, \varphi_{\2}(x) |\\
& = &
| \varphi_{\2}(x) | \, | \dot{\varphi}_{\1}(x) - \dot{\varphi}_{\2}(x)
\, \varphi_{\2}^{- \1}(x) \, \varphi_{\1}(x) |\\
& = &
| \dot{\varphi}_{\1}(x) | \, | \varphi_{\2}(x) - \varphi_{\1}(x)
\, \dot{\varphi}_{\1}^{- \1}(x) \, \dot{\varphi}_{\2}(x) |\\
& = &
| \dot{\varphi}_{\2}(x) | \, | \varphi_{\1}(x) - \varphi_{\2}(x)
\, \dot{\varphi}_{\2}^{- \1}(x) \, \dot{\varphi}_{\1}(x) |~.
\end{eqnarray*}
Solutions of equation (\ref{qlde}) 
\[ \varphi_{\1 , \2}(x) =  \exp[ \, q_{\1, \2} \, x \, ] = 
\exp[ \, ( \, p_{\1, \2} - \mbox{$\frac{a_{\0}}{2}$}  \,   
) \, x \, ]
\]
are given in terms of the solutions of the quadratic 
equation (\ref{qua2}), $p_{\1 ,\2}$,  and of the real variable 
$x$. In this case, the Wronskian becomes
\[ \mathcal{W}(x) = | p_{\1} - p_{\2} | \, | \exp [ q_{\1} x ] | \,
 | \exp [ q_{\2} x ]|~.\] 
This functional allows to distinguish between quaternionic linear 
dependent ($\mathcal{W}=0$) and independent ($\mathcal{W} \neq 0$)
solutions. A generalization for quaternionic
second order differential equations with non constant coefficients should be
investigated.

For $p_{\1} \neq p_{\2}$, the solution of equation (\ref{qlde}) is then
given by
\begin{equation}
\label{sol12}
\varphi(x) = \exp[ - \mbox{$\frac{a_{\0}}{2}$} \, x] \left\{  
\exp[ p_{\1} \, x ]  \, c_{\1} + \exp[ p_{\2} \, x ] \, c_{\2} \right\}~.
\end{equation}
As observed at the end of the previous subsection, the fundamental theorem of 
algebra is lost for a {\em restricted} class of quaternionic quadratic 
equation, i.e. 
$ p^{\2} + \alpha^{\2} = 0$ 
where $\alpha \in \mathbb{R}$. For these equations we  find  an infinite 
number of solutions, $p = \boldsymbol{h} \cdot \boldsymbol{\alpha}$ with  
$| \boldsymbol{\alpha}|^{\2} = \alpha^{\2}$. Nevertheless, the general 
solution of the second order differential equation 
\begin{equation}
\label{ed1}
\ddot{\varphi}(x) + \alpha^{\2} \, \varphi(x) = 0~,
\end{equation}
is also expressed in terms of  {\em two} linearly independent 
exponential solutions
\begin{equation}
\label{sol122}
\varphi(x) =   
\exp[ i \, \alpha  \, x ]  \, c_{\1} + 
\exp[ - i \, \alpha  \, x ] \, c_{\2}~.
\end{equation}
Note that any other exponential solution,   
$\exp [ \boldsymbol{h} \cdot \boldsymbol{\alpha} \, x ]$,   
can be written as  linear combination of  $\exp[i \, \alpha \, x]$ and
$ \exp[- i \, \alpha  \, x]$,
\[ 
\exp [ \boldsymbol{h} \cdot \boldsymbol{\alpha} \, x ] = 
\mbox{$\frac{1}{2 \alpha}$} \, \left\{
\exp[  i \, \alpha  \, x] \, \left(  \alpha  - i \, 
\boldsymbol{h} \cdot \boldsymbol{\alpha} \right) + 
\exp[  - i \, \alpha  \, x] \, \left(  \alpha  + i \, 
\boldsymbol{h} \cdot \boldsymbol{\alpha} \right) \right\}~.
\] 
As consequence, the loss of the fundamental theorem of algebra for 
quaternions does {\em not} represent an obstacle in solving second order 
quaternionic linear differential equations with constant coefficients.  
To complete our discussion,
we have to examine the case $p_{\1}=p_{\2}$. From equation 
(\ref{sc}) we find
\[ p_{\1} = p_{\2} =  -
\mbox{$\frac{ \boldsymbol{h} \times \boldsymbol{a}}{2}$} +  
\mbox{$\frac{1}{ |\boldsymbol{a}|^{\2}}$} \, 
 \boldsymbol{h} \cdot  \boldsymbol{a} \times \left(
 \boldsymbol{b} - \mbox{$\frac{ a_{\0}}{2}$} \, 
 \boldsymbol{a} \right)~,
\]
Thus, a first solution of the differential equation (\ref{qlde}) is 
\[
\xi(x) = 
\exp \left\{ \left[ \boldsymbol{h} \cdot \left( 
\mbox{$\frac{ \boldsymbol{a} \times \boldsymbol{b}}{|\boldsymbol{a}|^{\2}}$} - 
\mbox{$\frac{ \boldsymbol{a}}{2}$} \right)
-
\mbox{$\frac{a_{\0}}{2}$}   
 \right] x \right\}~.
\]
For  $\boldsymbol{a} \times \boldsymbol{b} = 0$, 
we can immediately obtain a second 
linearly independent solution by multiplying 
$\exp [ - \mbox{$\frac{a}{2}$} \, x]$ 
by $x$, $\eta(x) = x \, \xi(x)$. For  
$\boldsymbol{a} \times \boldsymbol{b} \neq 0$,
the second linearly independent solution takes a more complicated form, i.e.
\begin{equation}
\label{til}
\eta (x) = \left( x + \mbox{$\frac{ \boldsymbol{h} \cdot 
\boldsymbol{a}}{|\boldsymbol{a}|^{\2}}$} \right) \, \xi(x)~.
\end{equation}
It can easily be shown that $\eta(x)$ is solution of the differential 
equation (\ref{qlde}), 
\begin{eqnarray*}
\ddot{\eta}(x) + a \, 
\dot{\eta}(x) + b \, \eta(x) & = & 
\left[ x \, \left( q^{\2} + a \, q + b \right) + 2 \, q + a +
\mbox{$\frac{ \boldsymbol{h} \cdot 
\boldsymbol{a}}{|\boldsymbol{a}|^{\2}}$} \, \left( q^{\2} + a \, q \right)
+ b \, \mbox{$\frac{ \boldsymbol{h} \cdot 
\boldsymbol{a}}{|\boldsymbol{a}|^{\2}}$}
\right] \, 
\xi(x)\\
 & = &
\left( 2 \, q + a  + \left[ b \, , \, \mbox{$\frac{ \boldsymbol{h} \cdot 
\boldsymbol{a}}{|\boldsymbol{a}|^{\2}}$} \right]
\right) \, 
\xi(x)\\
 & = & 
\left( 2 \, \boldsymbol{h} \cdot  \mbox{$\frac{ \boldsymbol{a} \times 
\boldsymbol{b}}{|\boldsymbol{a}|^{\2}}$} +
 \left[ \boldsymbol{h} \cdot 
\boldsymbol{b} \, , \, \mbox{$\frac{ \boldsymbol{h} \cdot 
\boldsymbol{a}}{|\boldsymbol{a}|^{\2}}$} \right]
\right) \, \xi(x)\\
 & = & 0~.
\end{eqnarray*}
Thus, for $p_{\1}=p_{\2}= p 
=\boldsymbol{h} \cdot \left( \mbox{$\frac{ \boldsymbol{a} \times 
\boldsymbol{b}}{|\boldsymbol{a}|^{\2}}$} - 
\mbox{$\frac{ \boldsymbol{a}}{2}$} \right)$, the general solution
of the differential equation (\ref{qlde}) is given by  
\begin{equation}
\label{soli}
\varphi(x) = \exp[ - \mbox{$\frac{a_{\0}}{2}$} \, x] \left\{  
\exp[ p \, x ]  \, c_{\1} + \left( x + \mbox{$\frac{ \boldsymbol{h} \cdot 
\boldsymbol{a}}{|\boldsymbol{a}|^{\2}}$} \right) \,
\exp[ p \, x ] \, c_{\2} \right\}~.
\end{equation}


\subsection{Diagonalization and Jordan form}

To find the general solution of linear differential equations, we can also use
quaternionic formulations of 
eigenvalue equations, matrix diagonalization and Jordan form.
The quaternionic linear differential equation (\ref{qlde}) can be written 
in matrix form as follows 
\begin{equation}
\label{matrix}
\dot{\Phi} (x) = M \, \Phi (x)~,
\end{equation}
where
\[
M = \left( \begin{array}{cc} 0 & ~1\\ $-$b & ~$-$a \end{array} \right)~~~
\mbox{and}~~~
\Phi(x) = \left[ \begin{array}{c} \varphi (x)\\ 
\dot{\varphi} (x) \end{array} \right]~.
\]
By observing that $x$ is real, the formal solution of the matrix 
equation (\ref{matrix}) is given by
\begin{equation}
\Phi (x) = \exp [ M \, x ] \, \Phi(0)~,
\end{equation}
where $\Phi(0)$ represents a constant quaternionic column vector determined
by the initial conditions $\varphi(0)$, $\dot{\varphi}(0)$ and 
$\exp [ M \, x ] = \displaystyle{\sum_{n =0}^{\infty}} \frac{(Mx)^{n}}{n!}$.
In the sequel, we shall use right eigenvalue equations for quaternionic
linear matrix operators  
equations
\begin{equation}
\label{eig}
M \, \Phi = \Phi \, q~.
\end{equation}
Without loss of generality, we can work with 
{\em complex} eigenvalue equations. By setting $\Psi = \Phi u$, from the 
previous equation, we have 
\begin{equation}
\label{eigc}
M \, \Psi = M \, \Phi u = \Phi \, q \, u =   
\Phi u \, \overline{u} q  u = \Psi \, z~,
\end{equation}
where $z \in \mathbb{C}$ and $u$ is a unitary quaternion. 
In a recent paper~\cite{DEL00}, we find a complete discussion of the eigenvalue
equation for quaternionic matrix operators. In such a paper was shown that  
the complex counterpart of the matrix $M$ has an eigenvalue spectrum 
characterized by  eigenvalues which appear in conjugate pairs 
$\{ z_{\1} , \overline{z}_{\1}, z_{\2} , \overline{z}_{\2} \}$. 
Let $\Psi_{\1}$ and $\Psi_{\2}$ be the quaternionic 
eigenvectors corresponding to the complex eigenvalues $z_{\1}$ and $z_{\2}$
\[ M \, \Psi_{\1} = \Psi_{\1} \, z_{\1}~~~\mbox{and}~~~
M \, \Psi_{\2} = \Psi_{\2} \, z_{\2}~.\]
It can be shown that for $|z_{\1}| \neq |z_{\2}|$, the eigenvectors 
$\Psi_{\1}$ and $\Psi_{\2}$ are linearly independent on $\mathbb{H}$ and 
consequently there exists a $2 \times 2$ quaternionic matrix 
$S= \left[ \, \Psi_{\1} \, \Psi_{\2} \, \right]$
which diagonalizes $M$, 
\begin{eqnarray*}
\exp [ M \, x  ] = 
S \, \exp \left[  
\left( \begin{array}{cc} z_{\1}  & ~0\\ 
0 & ~z_{\2}  \end{array} \right) \, x  
\right] \, S^{\mi \1 } =  S \,   
\left( \begin{array}{cc}  \exp [ z_{\1} x] & 0\\ 
0 & \exp [ z_{\2} x] \end{array} \right) 
\, S^{\mi \1}~.
\end{eqnarray*}
In this case, the general solution of the quaternionic 
differential equation can be 
written in terms of the elements of the matrices $S$
and $S^{\mi \1}$ and of the complex eigenvalues $z_{\1}$ and $z_{\2}$, 
\begin{eqnarray*}
 \left[ \begin{array}{c} \varphi (x)\\ 
\dot{\varphi} (x) \end{array} \right]   =    
\left( \begin{array}{cc}  S_{\1 \1} \, \exp [ z_{\1} x] & 
~S_{\1 \2} \, \exp [ z_{\2} x] \\ 
S_{\2 \1} \, \exp [ z_{\1} x] & 
~S_{\2 \2} \, \exp [ z_{\2} x] \end{array} \right)
\,
\left[ \begin{array}{c} 
S_{\1 \1}^{\mi \1} \, \varphi(0) +
S_{\1 \2}^{\mi \1} \, \dot{\varphi}(0)\\
S_{\2 \1}^{\mi \1} \, \varphi(0) +
S_{\2 \2}^{\mi \1} \, \dot{\varphi}(0)
\end{array}
\right]
~.
\end{eqnarray*}
Hence, 
\begin{eqnarray}
\label{solm2}
\varphi (x) & = & S_{\1 \1} \, \exp[z_{\1} \, x] \, \left[  
S_{\1 \1}^{\mi \1} \, \varphi(0) + 
S_{\1 \2}^{\mi \1} \, \dot{\varphi}(0) \right] + \nonumber \\
 & &  
S_{\1 \2} \, \exp[z_{\2} \, x] \, \left[
S_{\2 \1}^{\mi \1} \, \varphi(0) +
S_{\2 \2}^{\mi \1} \, \dot{\varphi}(0) \right]
\nonumber \\
 & = & 
\exp \left[ S_{\1 \1} \, z_{\1} \,  \left( S_{\1 \1} \right)^{\mi \1} \, 
x \right] \, S_{\1 \1} \, \left[  
S_{\1 \1}^{\mi \1} \, \varphi(0) + 
S_{\1 \2}^{\mi \1} \, \dot{\varphi}(0) \right] + \nonumber \\
 & &  
\exp \left[ S_{\1 \2} \, z_{\2} \,  \left( S_{\1 \2} \right)^{\mi \1} \, 
x \right] \, S_{\1 \2} \, \left[
S_{\2 \1}^{\mi \1} \, \varphi(0) +
S_{\2 \2}^{\mi \1} \, \dot{\varphi}(0) \right] 
\nonumber \\
 & = &  
\exp \left[ S_{\2 \1} \, \left( S_{\1 \1} \right)^{\mi \1} \, 
x \right] \, 
S_{\1 \1} \, \left[  
S_{\1 \1}^{\mi \1} \, \varphi(0) + 
S_{\1 \2}^{\mi \1} \, \dot{\varphi}(0) \right] + \nonumber \\ 
& & 
\exp \left[ S_{\2 \2} \, \left( S_{\1 \2} \right)^{\mi \1} \, 
x \right] \, 
S_{\1 \2} \, \left[
S_{\2 \1}^{\mi \1} \, \varphi(0) +
S_{\2 \2}^{\mi \1} \, \dot{\varphi}(0) \right]~. 
\end{eqnarray}
We remark that a 
different choice of the eigenvalue spectrum does {\em not} modify the
solution (\ref{solm2}). In fact, by taking the following quaternionic 
eigenvalue 
spectrum 
\begin{equation}
\label{spe} 
\left\{ \, 
q_{\1} \, , \,  
q_{\2} \, \right\} =
\left\{ \, 
\overline{u}_{\1} z_{\1} u_{\1} \, , \,  
\overline{u}_{\2} z_{\2} u_{\2} \, \right\}~,~~~|q_{\1}| \neq |q_{\2}|~,
\end{equation}
and observing that the corresponding linearly independent 
eigenvectors are given by
\begin{equation}
\label{es}
\left\{ \, 
\Phi_{\1} = \Psi_{\1} u_{\1} \, , \,  
\Phi_{\2} = \Psi_{\2} u_{\2} \, \right\}~,
\end{equation}
we obtain
\begin{eqnarray*}
M & = &  
 \left[ \, \Phi_{\1}  ~ \Phi_{\2}  \, \right] \, 
\mbox{diag}
\left\{ 
q_{\1},   
q_{\2} \right\}
\,
\left[ \, \Phi_{\1}  ~ \Phi_{\2} \, \right]^{\mi \1}\\
 & = & 
\left[ \, \Psi_{\1}  u_{\1} ~ \Psi_{\2}  u_{\2}  \, \right] \, 
\mbox{diag}
\left\{ 
\overline{u}_{\1} z_{\1}  u_{\1} ,   
\overline{u}_{\2} z_{\2}  u_{\2} \right\}
\,
\left[ \, \Psi_{\1} u_{\1}  ~ \Psi_{\2} u_{\2} \, \right]^{\mi \1}\\
 & = &
 \left[ \, \Psi_{\1}  ~ \Psi_{\2}  \, \right] \, 
\mbox{diag}
\left\{ 
z_{\1},   
z_{\2} \right\}
\,
\left[ \, \Psi_{\1}  ~ \Psi_{\2} \, \right]^{\mi \1}~.
\end{eqnarray*}
Let us now discuss the case $|z_{\1}| = |z_{\2}|$. If the eigenvectors
$ \left\{ \, \Psi_{\a} \, , \,  \Psi_{\b}  \, \right\}$,   
corresponding to the eigenvalue spectrum 
$\left\{ \, z \, , \, z \, \right\}$,  
are linearly independent on $\mathbb{H}$, we can obviously repeat the previous
discussion and diagonalize the matrix operator $M$ by the $2 \times 2$ 
quaternionic matrix  $U = \left[ \, \Psi_{\1}  ~ \Psi_{\2}  \, \right]$. Then,
we find the 
\begin{eqnarray}
\label{solm22}
 \varphi (x) 
 & = & 
\exp \left[ U_{\1 \1} \, z \,  \left( U_{\1 \1} \right)^{\mi \1} \, 
x \right] \, U_{\1 \1} \, \left[  
U_{\1 \1}^{\mi \1} \, \varphi(0) + 
U_{\1 \2}^{\mi \1} \, \dot{\varphi}(0) \right] + \nonumber \\
 & &  
\exp \left[ U_{\1 \2} \, z \,  \left( U_{\1 \2} \right)^{\mi \1} \, 
x \right] \, U_{\1 \2} \, \left[
U_{\2 \1}^{\mi \1} \, \varphi(0) +
U_{\2 \2}^{\mi \1} \, \dot{\varphi}(0) \right] 
\nonumber \\
  & = &  
\exp \left[ U_{\2 \1} \, \left( U_{\1 \1} \right)^{\mi \1} \, 
x \right] \, 
U_{\1 \1} \, \left[  
U_{\1 \1}^{\mi \1} \, \varphi(0) + 
U_{\1 \2}^{\mi \1} \, \dot{\varphi}(0) \right] + \nonumber \\ 
& & 
\exp \left[ U_{\2 \2} \, \left( U_{\1 \2} \right)^{\mi \1} \, 
x \right] \, 
U_{\1 \2} \, \left[
U_{\2 \1}^{\mi \1} \, \varphi(0) +
U_{\2 \2}^{\mi \1} \, \dot{\varphi}(0) \right]~. 
\end{eqnarray}
For
linearly dependent eigenvectors, we cannot construct
a matrix which diagonalizes the matrix operator $M$. Nevertheless, 
we can transform the matrix operator $M$ to Jordan form  
\begin{equation}
\label{jf}
M = J \, \left( 
\begin{array}{cc} 
z & ~1\\ 0 & ~z 
\end{array}
\right)
\, J^{\mi \1}~.
\end{equation}
It follows that the solution of our quaternionic 
differential equation can be written as
\begin{eqnarray*}
\label{solm3}
\Phi(x) & = &J \, \exp \left[     \left( 
\begin{array}{cc} 
z & ~1\\ 0 & ~z 
\end{array}
\right) \,  x     \right] \, 
J^{\mi \1} \, \Phi(0) \nonumber \\
 & = & 
\left( 
\begin{array}{cr} 
J_{\1 \1}  & ~ x \, J_{\1 \1} + J_{\1 \2} \\
J_{\2 \1}  & ~ x \, J_{\2 \1} + J_{\2 \2} 
\end{array}
\right) \, \exp[zx] \, 
\left[ \begin{array}{c} 
J_{\1 \1}^{\mi \1} \, \varphi(0) +
J_{\1 \2}^{\mi \1} \, \dot{\varphi}(0)\\
J_{\2 \1}^{\mi \1} \, \varphi(0) +
J_{\2 \2}^{\mi \1} \, \dot{\varphi}(0)
\end{array}
\right]~.
\end{eqnarray*}
Thus,
\begin{eqnarray}
\label{solm4}
\varphi (x) & = & J_{\1 \1} \, \exp[z \, x] \, \left[  
J_{\1 \1}^{\mi \1} \, \varphi(0) + 
J_{\1 \2}^{\mi \1} \, \dot{\varphi}(0) \right] + \nonumber \\
 & &  
\left( x \, J_{\1 \1} + J_{\1 \2} \right) \, \exp[z \, x] \, \left[
J_{\2 \1}^{\mi \1} \, \varphi(0) +
J_{\2 \2}^{\mi \1} \, \dot{\varphi}(0) \right]
\nonumber \\
 & = & 
\exp \left[ J_{\1 \1} \, z  \,  \left( J_{\1 \1} \right)^{\mi \1} \, 
x \right] \, J_{\1 \1} \, \left[  
J_{\1 \1}^{\mi \1} \, \varphi(0) + 
J_{\1 \2}^{\mi \1} \, \dot{\varphi}(0) \right] + \nonumber \\
 & &  
\left[ x  + J_{\1 \2}  \left( J_{\1 \1} \right)^{\mi \1} \right] \,
\exp \left[ J_{\1 \1} \, z  \,  \left( J_{\1 \1} \right)^{\mi \1} \, 
x \right] \times \nonumber \\
 & &
J_{\1 \1} \,
\left[
J_{\2 \1}^{\mi \1} \, \varphi(0) +
J_{\2 \2}^{\mi \1} \, \dot{\varphi}(0) \right] 
\nonumber \\
 & = &  
\exp \left[ J_{\2 \1} \, \left( J_{\1 \1} \right)^{\mi \1} \, 
x \right] \, 
J_{\1 \1} \, \left[  
J_{\1 \1}^{\mi \1} \, \varphi(0) + 
J_{\1 \2}^{\mi \1} \, \dot{\varphi}(0) \right] + \nonumber \\ 
& & 
\left[ x  + J_{\1 \2}  \left( J_{\1 \1} \right)^{\mi \1} \right] \,
\exp \left[ J_{\2 \1} \,  \left( J_{\1 \1} \right)^{\mi \1} \, 
x \right] \times \nonumber \\
 & &
J_{\1 \1} \,
\left[
J_{\2 \1}^{\mi \1} \, \varphi(0) +
J_{\2 \2}^{\mi \1} \, \dot{\varphi}(0) \right]~. 
\end{eqnarray}
The last equality in the previous equation follows from the use of equation
(\ref{jf}) and the definition of $M$.

Finally, the general solution of the quaternionic differential equation
(\ref{qlde}) can be given by solving the corresponding eigenvalue problem.
We conclude this section, by observing that the quaternionic exponential
solution, $\exp[q \, x]$,  can also be written in terms 
of complex exponential solutions, 
$u \, \exp[z \, x] \, u^{\mi \1}$, where $q = u \, z \, u^{\mi \1}$. 
 The elements of  the similarity transformations $S$, $U$  or $J$
and the complex eigenvalue spectrum of $M$ determine the 
quaternion  $u$ and the  complex number $z$. 
This form for exponential solutions 
will be very useful in solving complex linear differential equations with 
constant coefficients. In fact, due to the presence of the right acting 
operator $R_{\i}$, we cannot use quaternionic exponential solutions
for complex linear differential equations.


\section{COMPLEX LINEAR QUATERNIONIC DIFFERENTIAL EQUATIONS}
\label{sec7}

Consider now the second order complex linear quaternionic 
differential operator
\begin{eqnarray*}
\mathcal{D}_{\Co} & = &
\left[   
a_{\0 \2} + \boldsymbol{L} \cdot \boldsymbol{a}_{\2} + 
\left( 
b_{\0 \2 } + \boldsymbol{L} \cdot \boldsymbol{b}_{\2}
\right) \, R_{\i} \right] \, 
 \partial_{\xx} + \\
& & 
\left[   
a_{\0 \1} + \boldsymbol{L} \cdot \boldsymbol{a}_{\1} + 
\left( 
b_{\0 \1 } + \boldsymbol{L} \cdot \boldsymbol{b}_{\1}
\right) \, R_{\i} \right] \, 
\partial_{\x} +\\
 & & ~ \, a_{\0 \0} + \boldsymbol{L} \cdot \boldsymbol{a}_{\0} + 
\left( 
b_{\0 \0 } + \boldsymbol{L} \cdot \boldsymbol{b}_{\0}
\right) \, R_{\i}\\
& \in & 
\mathcal{A_{\Co}} \otimes \mathcal{O}
\end{eqnarray*}
and look for solutions of the complex linear quaternionic 
differential equation
\begin{equation}
\label{clde}
\mathcal{D}_{\Co} \, \varphi(x) = 0~.
\end{equation} 
Due to the presence of $R_{i}$ in  (\ref{clde}),
the general solution of  the complex linear quaternionic 
differential equation cannot be given in terms of quaternionic
exponentials. In matrix form,  equation  (\ref{clde})  reads
\begin{equation}
\label{matrix2}
\dot{\Phi} (x) = M_{\Co} \, \Phi (x)~,
\end{equation}
where
\[
M_{\Co} = \left( \begin{array}{cc} 0 & ~1\\ $-$b_{\Co} & ~$-$a_{\Co} 
\end{array} \right)~~~
\mbox{and}~~~
\Phi(x) = \left[ \begin{array}{c} \varphi (x)\\ 
\dot{\varphi} (x) \end{array} \right]~.
\]
The complex counterpart of complex 
linear quaternionic matrix operator $M_{\Co}$ has an eigenvalue 
spectrum  characterized by four complex eigenvalues  
$\{ z_{\1} , z_{\2}, z_{\3} , z_{\4} \}$. It can be shown that 
$M_{\Co}$ is diagonalizable if and only if its complex counterpart is 
diagonalizable. For diagonalizable 
matrix operator $M_{\Co}$, we can find a complex linear quaternionic linear
similarity transformation $S_{\Co}$ which reduces the matrix operator 
$M_{\Co}$ to diagonal form~\cite{DEL00}
\[
M_{\Co} = S_{\Co} \, \left( \begin{array}{cc} 
\mbox{$\frac{z_{\1} + \overline{z}_{\2}}{2} + \frac{z_{\1} - \overline{z}_{\2}}{2i} \,
R_{\i}$} & 0 \\  
0 &  
\mbox{$\frac{z_{\3} + \overline{z}_{\4}}{2} + \frac{z_{\3} - \overline{z}_{\4}}{2i} \,
R_{\i}$} \end{array} \right)
\, S_{\Co}^{\mi \1}~.
\]
It is immediate to verify that 
\[
\left\{~\left( \begin{array}{c} 1 \\ 0\end{array} \right)~,~
  \left( \begin{array}{c} j \\ 0\end{array} \right)~,~
  \left( \begin{array}{c} 0 \\ 1\end{array} \right)~,~
  \left( \begin{array}{c} 0 \\ j\end{array} \right)~
\right\}  
 \]
are eigenvectors of the diagonal matrix operator
\[
\left( \begin{array}{cc} 
\mbox{$\frac{z_{\1} + \overline{z}_{\2}}{2} + \frac{z_{\1} - \overline{z}_{\2}}{2i} \,
R_{\i}$} & 0 \\  
0 &  
\mbox{$\frac{z_{\3} + \overline{z}_{\4}}{2} + \frac{z_{\3} - \overline{z}_{\4}}{2i} \,
R_{\i}$} \end{array} \right)
\]
with right complex eigenvalues $z_{\1}$, $z_{\2}$, $z_{\3}$ and  $z_{\4}$. 
The general solution  of the differential equation  (\ref{clde}) 
can be given in terms of these complex eigenvalues, 
\begin{eqnarray}
\label{solm2c}
\varphi (x) & = & S_{\Co \, \1 \1} \, 
\exp \left[ \left( \mbox{$\frac{z_{\1} + \overline{z}_{\2}}{2} + \frac{z_{\1} - 
\overline{z}_{\2}}{2i} \,
R_{\i}$} \right) \, x \right] \, \left[  
S_{\Co \, \1 \1}^{\mi \1} \, \varphi(0) + 
S_{\Co \, \1 \2}^{\mi \1} \, \dot{\varphi}(0) \right] + \nonumber \\
 & &  
S_{\Co \, \1 \2} \, \exp \left[ \left( 
\mbox{$\frac{z_{\3} + \overline{z}_{\4}}{2} + \frac{z_{\3} - \overline{z}_{\4}}{2i} \,
R_{\i}$} \right)\, x \right]
 \, \left[
S_{\Co \, \2 \1}^{\mi \1} \, \varphi(0) +
S_{\Co \, \2 \2}^{\mi \1} \, \dot{\varphi}(0) \right]
\nonumber \\
 & = & 
 u_{\1} \,  \exp[z_{\1} \, x] \, k_{\1} + 
u_{\2} \,  \exp[z_{\2} \, x] \, k_{\2} + \nonumber \\
 & &  u_{\3} \,  \exp[z_{\3} \, x] \, k_{\3} + 
u_{\4} \,  \exp[z_{\4} \, x] \, k_{\4}~,  
\end{eqnarray}
where $k_{\n}$ are complex coefficients determined by the 
initial conditions. This solution holds for diagonalizable
matrix operator $M_{\Co}$. For non diagonalizable 
matrix operators we need to find the similarity transformation
$J_{\Co}$ which reduces  $M_{\Co}$ to the  Jordan form. 
For instance, it can be shown that for equal eigenvalues, $z_{\1}=z_{\2}$,  
the general solution of the differential equation  (\ref{clde}) 
is
\begin{equation}
\label{cc22} 
 \varphi (x) =  u  \,  \exp[z  \, x] \, k_{\1} + 
\left( u \, x + \tilde{u} \right)  \,  \exp[z  \, x] \, k_{\2}
+  u_{\3} \,  \exp[z_{\3} \, x] \, k_{\3} + 
u_{\4} \,  \exp[z_{\4} \, x] \, k_{\4}~.
\end{equation}


\subsection{Schr\"odinger equation}

Let us now examine the complex linear 
Schr\"odinger equation in presence of a constant quaternionic potential,
\begin{equation}
\label{s2}
\left[ \mbox{$\frac{\hbar^{\2}}{2m}$} \, \partial_{\xx} 
- V   + j \, W  \right] \Psi(x)   
= i \, \Psi(x)  \, i \, E~.
\end{equation}
In this case, the complex linear matrix operator 
\[
M_{\Co} = \left( \begin{array}{cc} 0 & ~1\\ 
$-$ b_{\Co}   & ~0 
\end{array} \right)~,~~~b_{\Co} = V   - j \, W  + i \, E \, R_{\i}~,
\]
represents a diagonalizable operator. Consequently, the general
solution of the Schr\"odinger equation is given by 
\begin{equation}
\label{cc223} 
 \varphi (x) =  u_{\1}  \,  \exp[z_{\1}  \, x] \, k_{\1} + 
u_{\2} \, \exp[z_{\2}  \, x] \, k_{\2}
+  u_{\3} \,  \exp[z_{\3} \, x] \, k_{\3} + 
u_{\4} \,  \exp[z_{\4} \, x] \, k_{\4}~.
\end{equation}
The quaternions  $u_{\n}$ and the complex eigenvalues $z_{\n}$
are obtained by solving the eigenvalue equation  for
the complex linear operator $M_{\Co}$. We can also 
obtain the general solution of equation (\ref{s2}) by 
substituting $u \, 
\exp[\mbox{\footnotesize $\sqrt{\frac{2m}{\hbar^{\2}}}$} \, z \, x]$ in
the Schr\"odinger equation. We find the following quaternionic
equation
\[
 u \, z^{\2} - 
\left( V - j \, W \right) u - i \, E \,  u \, i = 0~,
\]  
where $u =  z_{\u} + j \, \tilde{z}_{\u}$. 
This equation can be written as two complex equations
\begin{equation*}
\left[ z^{\2} - (V - E) \right] z_{\u} - \overline{W} \tilde{z}_{\u} = 
\left[ z^{\2} - (V + E) \right] \tilde{z}_{\u} +  W \, z_{\u} = 0~.  
\end{equation*}
An easy calculation shows that $z$ satisfies the complex equation
\begin{equation}
\label{z4}
z^{\4} - 2 \, V \, z^{\2} + V^{\2}  + |W|^{\2} -  E^{\2} = 0~,
\end{equation}
whose roots are
\begin{equation}
 z_{\1 , \2} = \pm \, \mbox{\footnotesize $\sqrt{ V - \sqrt{E^{\2} - |W|^{\2}}}$}
= \pm \, z_{\mi} 
~~~\mbox{and}~~~
z_{\3 , \4} = \pm \, \mbox{\footnotesize $\sqrt{V + \sqrt{E^{\2} - |W|^{\2}}}$}
= \pm \, z_{\pl}~.
\end{equation}
By setting $\left( u_{\1 , \2} \right)_{\Co} = 
\left( - j  u_{\3 , \4} \right)_{\Co}= 1$, we find
\begin{equation}
\label{u}
u_{\mi} = \mbox{\footnotesize 
$\left( 1 + j \, \frac{W}{E + \sqrt{E^{\2} - |W|^{\2}}}
\right)$}~~~\mbox{and}~~~
u_{\pl} = \mbox{\footnotesize $
\left(\frac{\overline{W}}{E + \sqrt{E^{\2} - |W|^{\2}}} + 
j \right)$}~.
\end{equation}
The solution of the complex linear quaternionic Schr\"odinger equation
is then given by
\begin{eqnarray}
\label{slos}
\Psi(x) & = &   u_{\mi} \,  \left\{ \exp[ 
\mbox{\footnotesize $
\sqrt{\frac{2m}{\hbar^{\2}}}$} \, z_{\mi}  \, x] \, k_{\1} + 
\exp[- \, \mbox{\footnotesize $
\sqrt{\frac{2m}{\hbar^{\2}}}$} \, z_{\mi}  \, x] \, k_{\2} \right\} +
\nonumber \\ 
 & &  
u_{\pl} \,  \left\{ \exp[ \mbox{\footnotesize $
\sqrt{\frac{2m}{\hbar^{\2}}}$} \, z_{\pl} \, x] \, 
k_{\3} + 
\exp[- \,  \mbox{\footnotesize $
\sqrt{\frac{2m}{\hbar^{\2}}}$} \, z_{\pl} \, x] \, k_{\4} \right\}~.
\end{eqnarray}
Equation (\ref{z4}) can also be obtained by 
multiplying  the complex linear Schr\"odinger equation (\ref{s2}) 
from the left
by the operator
\[
\mbox{$\frac{\hbar^{\2}}{2m}$} \, \partial_{\xx} 
- V   - j \, W ~.   
\]
This gives\\

\noindent
\begin{tabular}{lcl}
$\left[ \left( \mbox{$\frac{\hbar^{\2}}{2m}$} \right)^{\2} \, \partial_{\xxxx} 
\mbox{$-$} 2 \, \mbox{$\frac{\hbar^{\2}}{2m}$}  \, V \,  \partial_{\xx} 
\mbox{$+$} V^{\2}   + | W |^{\2}   \right] \Psi(x)$ & $=$ &   $i \, \left[  
\mbox{$\frac{\hbar^{\2}}{2m}$} \, \partial_{\xx} 
- V   + j \, W  \right] \Psi(x)  \, i \, E$\\ 
&  $=$ &  $  E^{\2} \, \Psi(x)  $~.
\end{tabular}

\vspace*{0.3cm}

\noindent By substituting the exponential solution $u \, 
\exp[\mbox{\footnotesize 
$\sqrt{\frac{2m}{\hbar^{\2}}}$} \, z \, x]$ in the previous equation, 
we immediately re-obtain equation (\ref{z4}).


\section{QUATERNIONIC CONSTANT POTENTIALS}
\label{sec8}

Of all Schr\"odinger equations the one for a constant potential is 
mathematically the simplest. The reason for resuming the study of the
Schr\"odinger equation with such a potential is that the qualitative
features of a physical potential can often be approximated reasonably well by
a potential which is pieced together from a number of constant portions.


\subsection{The potential step}
\label{ss61}

Let us consider the quaternionic potential step,
\[
V(x) - j \, W(x) = \left\{ \begin{array}{cl}
                                       0 & ~~x<0\\
                         V - j \, W & ~~x>0  
               \end{array}  \right. ~,
\]
where $V$ and $W$ represent constant potentials. 
For scattering problems with a wave function incident from the left 
on the quaternionic potential step, the complex linear
quaternionic Schr\"odinger equation has solution
\begin{equation}
 \Psi(x) = \left\{ \begin{array}{ll}
x<0~: & \\
\exp [ \,  i \,  
\mbox{$\frac{p}{\hbar}$} \, x \, ] + r \, \exp[ \, - i \,  
\mbox{$\frac{p}{\hbar}$} \, x \, ] + 
j \, \tilde{r} \, \exp[ \,  
\mbox{$\frac{p}{\hbar}$} \, x \,]~; & \\
 & \\
x>0~: & \\
u_{\mi} \, t \, \exp[ 
\mbox{\footnotesize $
\sqrt{\frac{2m}{\hbar^{\2}}}$} \, z_{\mi}  \, x]  +
u_{\pl} \,  \tilde{t} \, 
\exp[- \,  \mbox{\footnotesize $
\sqrt{\frac{2m}{\hbar^{\2}}}$} \, z_{\pl} \, x]  
& ~~\mbox{\footnotesize $[\, E > \sqrt{V^{\2}+ |W|^{\2}} \,]$}~,\\
u_{\mi} \, t \, \exp[ - 
\mbox{\footnotesize $
\sqrt{\frac{2m}{\hbar^{\2}}}$} \, z_{\mi}  \, x] +
u_{\pl} \,  \tilde{t} \, 
\exp[- \,  \mbox{\footnotesize $
\sqrt{\frac{2m}{\hbar^{\2}}}$} \, z_{\pl} \, x]  
& ~~\mbox{\footnotesize $[\, E < \sqrt{V^{\2}+ |W|^{\2}} \,]$}~. 
                \end{array}  \right. 
\end{equation}
where $r$, $\tilde{r}$, $t$ and $\tilde{t}$ are complex coefficients to be 
determined by matching the wave function $\Psi(x)$ and its slope at the 
discontinuity of the potential $x=0$.

For $E > \sqrt{V^{\2}+ |W|^{\2}}$, 
the complex exponential solutions of the quaternionic Schr\"odinger equation  
are characterized by 
\[
 z_{\mi} = i \, \mbox{\footnotesize $\sqrt{ \sqrt{E^{\2} - |W|^{\2}} - V}$}
~\in ~i \, \mathbb{R}~~~~~
\mbox{and}~~~~~
z_{\pl} = \mbox{\footnotesize $\sqrt{\sqrt{E^{\2} - |W|^{\2}} +V}$} ~\in~ 
\mathbb{R} ~.
\]
The complex linearly independent solutions 
\[ u_{\mi} \,  \exp[ - 
\mbox{\footnotesize $
\sqrt{\frac{2m}{\hbar^{\2}}}$} \, z_{\mi}  \, x]
~~~\mbox{and}~~~ 
u_{\pl} \, \exp[ \mbox{\footnotesize $
\sqrt{\frac{2m}{\hbar^{\2}}}$} \, z_{\pl} \, x]
\]
have been omitted, $k_{\2} = k_{\3} = 0$ in (\ref{slos}),  
because we are considering a wave incident from the left and 
because the second complex exponential solution,  $\exp[ \mbox{\footnotesize$
\sqrt{\frac{2m}{\hbar^{\2}}}$} \, z_{\pl} \, x]$,
is in conflict with the boundary condition that $\Psi(x)$ remain finite
as $x \to \infty$. The standard result of complex quantum mechanics are 
immediately recovered by considering $W=0$ and taking 
the complex part of the quaternionic
solution. 

For $E < \sqrt{V^{\2}+ |W|^{\2}}$,   
the complex exponential solutions of the quaternionic Schr\"odinger equation  
are characterized by \\

\noindent
\begin{tabular}{llll}
$z_{\mi} = \mbox{\footnotesize $\sqrt{ V - \sqrt{E^{\2} - |W|^{\2}}}$}~,~$
& $z_{\pl} = \mbox{\footnotesize $\sqrt{V + \sqrt{E^{\2} - |W|^{\2}}}$}$ 
& $~\in~ 
\mathbb{R} $
& $~~\mbox{\footnotesize $[ \, E> |W| \, ]$}~,$\\
$z_{\pem} = \mbox{\footnotesize $\left( V^{\2} + |W|^{\2} - E^{\2} 
\right)^{\fq}$}
 \,
\exp[ \pm \, i \, \mbox{$\frac{\theta}{2}$}]~,$
& $\tan \theta = \mbox{\footnotesize 
$ \frac{\sqrt{|W|^{\2} - E^{\2}}}{V}$} $
& $~\in~ 
\mathbb{C}$
& $~~\mbox{\footnotesize $[ \, E< |W| \,]$}~.$
\end{tabular}

\vspace*{0.3cm}

\noindent 
The complex linearly independent solutions 
\[ u_{\mi} \,  \exp[  
\mbox{\footnotesize $
\sqrt{\frac{2m}{\hbar^{\2}}}$} \, z_{\mi}  \, x]
~~~\mbox{and}~~~ 
u_{\pl} \, \exp[ \mbox{\footnotesize $
\sqrt{\frac{2m}{\hbar^{\2}}}$} \, z_{\pl} \, x]
\]
have been omitted, $k_{\1} = k_{\3} = 0$ in (\ref{slos}),  
because they are  in conflict with the boundary condition that 
$\Psi(x)$ remain finite as $x \to \infty$.

A relation between the complex coefficients of reflection and transmission
can immediately be obtained by the continuity equation
\begin{equation}
\label{cont}
\partial_{\t} \rho (x,t) + \partial_{\x} J (x,t) = 0~,
\end{equation}
where
\[ \rho(x,t) = \overline{\Phi}(x,t) \, \Phi(x,t)~,\]
and
\[
J(x,t) = 
\mbox{\footnotesize $\frac{\hbar}{2m}$} \, \left\{
\,
\left[ \partial_{\x} \overline{\Phi}(x,t) \right] \, i \, \Phi(x,t) -
\overline{\Phi}(x,t) \, i \, 
\partial_{\x} \Phi(x,t)  \, \right\}~.
\]
Note that, due to the non commutative nature of the quaternionic wave 
functions, the position of the imaginary unit $i$ in the probability current
density $J(x,t)$ is important to recover a continuity equation in quaternionic
quantum mechanics. For stationary states, 
$\Phi(x,t) = \Psi(x) 
\exp [ \, - i \, \mbox{$\frac{E}{\hbar}$} \, t \,] \zeta(0)$, it can easily be 
shown that the probability current density 
\[ 
 J(x,t) =
\mbox{\footnotesize $\frac{\hbar}{2m}$} \, \overline{\zeta}(0) \, 
\exp [ \, i \, \mbox{$\frac{E}{\hbar}$} \, t \,] 
\left\{
\,
\left[ \partial_{\x} \overline{\Psi}(x) \right] \, i \, \Psi(x) -
\overline{\Psi}(x) \, i \, 
\partial_{\x} \Psi(x)  \, \right\}
\, \exp [ \, - i \, \mbox{$\frac{E}{\hbar}$} \, t \,] \, \zeta(0)~.
\]
must be independent of $x$, $J(x,t) = f(t)$. Hence,
\[ 
 \mbox{\footnotesize $\frac{\hbar}{2m}$} \,
\left\{
\,
\left[ \partial_{\x} \overline{\Psi}(x) \right] \, i \, \Psi(x) -
\overline{\Psi}(x) \, i \, 
\partial_{\x} \Psi(x) \, \right\} = 
\exp [ \, - i \, \mbox{$\frac{E}{\hbar}$} \, t \,] \, \zeta(0)
\, f(t) \,  \overline{\zeta}(0) \, 
\exp [ \, i \, \mbox{$\frac{E}{\hbar}$} \, t \,] = \alpha~,
\] 
where $\alpha$ is a real constant. This implies that the quantity
\[
\mathcal{J} = \mbox{\footnotesize $\frac{p}{2m}$} \, \left\{ \, 
 \left[ \partial_{\x} \overline{\Psi}(x) \right] \, i \, \Psi(x) -
\overline{\Psi}(x) \, i \, 
\partial_{\x} \Psi(x)  \, \right\} 
\]
has the same value at all points $x$. In the free potential region, $x<0$,
we find
\begin{equation*}
\mathcal{J}_{\mi} = \mbox{\footnotesize $\frac{p}{m}$} \,
\left( \, 1 - |r|^{\2} \, \right)~.
\end{equation*}
In the potential region, $x>0$, we obtain
\begin{equation*}
 \mathcal{J}_{\pl} = \left\{ \begin{array}{ll}
\mbox{\footnotesize 
$\sqrt{ \mbox{\footnotesize $\frac{2}{m}$} \, \left( 
 \sqrt{E^{\2} - |W|^{\2}} - V \right)}$} \, \left[
1 - \left( 
\frac{|W|}{E + \sqrt{E^{\2} - |W|^{\2}}} \right)^{\2} \right] \,
|t|^{\2}
& ~~\mbox{\footnotesize $[\, E > \sqrt{V^{\2}+ |W|^{\2}} \,]$}~,\\
0 
& ~~\mbox{\footnotesize $[\, E < \sqrt{V^{\2}+ |W|^{\2}} \,]$}~. 
                \end{array}  \right. 
\end{equation*}
Finally, for stationary
states, the continuity equation leads to
\begin{equation}
\label{relation}
\begin{array}{ll}
|r|^{\2} + 
\mbox{ 
$\frac{\sqrt{E^{\2} - |W|^{\2}} - V}{E}$} \, \left[
1 - \left( 
\frac{|W|}{E + \sqrt{E^{\2} - |W|^{\2}}} \right)^{\2} \right] \,
|t|^{\2} = 1 
& ~~\mbox{\footnotesize $[\, E > \sqrt{V^{\2}+ |W|^{\2}} \,]$}~,\\
|r|^{\2} = 1 
& ~~\mbox{\footnotesize $[\, E < \sqrt{V^{\2}+ |W|^{\2}} \,]$}~. 
                \end{array}   
\end{equation} 
Thus, by using the concept of a probability current, we can define   
the following  coefficients of transmission and reflection 
\begin{equation*}
\begin{array}{lll}
R = |r|^{\2}~, &~T =  
\mbox{ 
$\frac{\sqrt{E^{\2} - |W|^{\2}} - V}{E}$} \, \left[
1 - \left( 
\frac{|W|}{E + \sqrt{E^{\2} - |W|^{\2}}} \right)^{\2} \right] \,
|t|^{\2}
& ~~\mbox{\footnotesize $[\, E > \sqrt{V^{\2}+ |W|^{\2}} \,]$}~,\\
R = |r|^{\2}~, &~ T=0~
& ~~\mbox{\footnotesize $[\, E < \sqrt{V^{\2}+ |W|^{\2}} \,]$}~. 
                \end{array}   
\end{equation*} 
These coefficients give the probability for the particle, arriving from 
$x=- \infty$, to pass the potential step at $x=0$ or to turn back. 
The coefficients $R$ and $T$ depend only on the ratios $E/V$ and
$|W|/V$. The predictions of complex quantum mechanics are recovered by 
setting $W=0$.


\subsection{The rectangular potential barrier}
\label{ss62}

In our study of quaternionic potentials, we now reach the rectangular 
potential barrier,
\[
V(x) - j \, W(x) = \left\{ \begin{array}{cl}
                                       0 & ~~x<0\\
                         V - j \, W & ~~0<x<a\\
                                       0 & ~~x>a  
               \end{array}  \right. ~.
\] 
For scattering problems with a wave function incident from the left 
on the quaternionic potential barrier, the complex linear
quaternionic Schr\"odinger equation has solution
\begin{equation}
 \Psi(x) = \left\{ \begin{array}{l}
x<0~:  \\
\exp [ \,  i \,  
\mbox{$\frac{p}{\hbar}$} \, x \, ] + r \, \exp[ \, - i \,  
\mbox{$\frac{p}{\hbar}$} \, x \, ] + 
j \, \tilde{r} \, \exp[ \,  
\mbox{$\frac{p}{\hbar}$} \, x \,]~; \\
 \\
0 <x < a~:  \\
u_{\mi} \,  \left\{ \exp[ 
\mbox{\footnotesize $
\sqrt{\frac{2m}{\hbar^{\2}}}$} \, z_{\mi}  \, x] \, k_{\1} + 
\exp[- \, \mbox{\footnotesize $
\sqrt{\frac{2m}{\hbar^{\2}}}$} \, z_{\mi}  \, x] \, k_{\2} \right\} +\\
u_{\pl} \,  \left\{ \exp[ \mbox{\footnotesize $
\sqrt{\frac{2m}{\hbar^{\2}}}$} \, z_{\pl} \, x] \, 
k_{\3} + 
\exp[- \,  \mbox{\footnotesize $
\sqrt{\frac{2m}{\hbar^{\2}}}$} \, z_{\pl} \, x] \, k_{\4} \right\}~;\\
 \\
x>a~:  \\
t \, \exp[ \, i \,  
\mbox{$\frac{p}{\hbar}$} \, x \, ] + 
j \, \tilde{t} \, \exp[ \, - \,   
\mbox{$\frac{p}{\hbar}$} \, x \,]~. \\
\end{array}  
\right. 
\end{equation}
The complex coefficients $r$, $\tilde{r}$, $t$ and $\tilde{t}$ are 
determined by matching the wave function $\Psi(x)$ and its slope at the 
discontinuity of the potential $x=0$ and will depend on $|W|$.

By using the continuity equation, we immediately find the following
relation between the transmission, $T=|t|^{\2}$, and reflection,
$R=|r|^{\2}$, coefficients
\begin{equation}
R + T = 1~.
\end{equation}


\subsection{The rectangular potential well}
\label{ss622}

Finally, we briefly discuss the quaternionic rectangular potential well,
\[
V(x) - j \, W(x) = \left\{ \begin{array}{cl}
                                       0 & ~~x<0\\
                         - V + j \, W & ~~0<x<a\\
                                       0 & ~~x>a  
               \end{array}  \right. ~.
\] 
In the potential region, the solution of the complex linear quaternionic 
Schr\"odinger equation is then given by
\begin{eqnarray}
\label{slos2}
\Psi(x) & = &   u_{\mi} \,  \left\{ \exp[ 
\mbox{\footnotesize $
\sqrt{\frac{2m}{\hbar^{\2}}}$} \, z_{\mi}  \, x] \, k_{\1} + 
\exp[- \, \mbox{\footnotesize $
\sqrt{\frac{2m}{\hbar^{\2}}}$} \, z_{\mi}  \, x] \, k_{\2} \right\} +
\nonumber \\
& & 
u_{\pl} \,  \left\{ \exp[ \mbox{\footnotesize $
\sqrt{\frac{2m}{\hbar^{\2}}}$} \, z_{\pl} \, x] \, 
k_{\3} + 
\exp[- \,  \mbox{\footnotesize $
\sqrt{\frac{2m}{\hbar^{\2}}}$} \, z_{\pl} \, x] \, k_{\4} \right\}~,
\end{eqnarray}
where
\begin{equation*}
u_{\mi} = \mbox{\footnotesize 
$\left( 1 - j \, \frac{W}{E + \sqrt{E^{\2} - |W|^{\2}}}
\right)$}~,~~~
u_{\pl} = \mbox{\footnotesize $
\left(j - \frac{\overline{W}}{E + \sqrt{E^{\2} - |W|^{\2}}} \right)$}~.
\end{equation*}
and 
\begin{equation*}
 z_{\mi} = i \, \mbox{\footnotesize $\sqrt{ \sqrt{E^{\2} - |W|^{\2}} + V }$} 
~,~~~
z_{\pl} = \mbox{\footnotesize $\sqrt{\sqrt{E^{\2} - |W|^{\2}} - V}$}
~.
\end{equation*}
Depending on whether the energy is positive or negative, we distinguish
two separate cases. If $E>0$, the particle is unconfined and is scattered
by the potential; if $E<0$, it is confined and in a bound state. 
We limit ourselves to discussing the case $E<0$. For
$ |W| < |E| < \sqrt{V^{\2} + |W|^{\2}}$, solution (\ref{slos2})
becomes
\begin{equation}
\label{slos22}
\begin{array}{l}
u_{\mi} \,  \left\{ \exp[ i \,  
\mbox{\footnotesize $
\sqrt{\frac{2m}{\hbar^{\2}}}$} \, 
\mbox{\footnotesize $\sqrt{ \sqrt{E^{\2} - |W|^{\2}} + V }$}   \, x] \, k_{\1} + 
\exp[- \, i \, \mbox{\footnotesize $
\sqrt{\frac{2m}{\hbar^{\2}}}$} \, 
\mbox{\footnotesize $\sqrt{ \sqrt{E^{\2} - |W|^{\2}} + V }$} 
  \, x] \, k_{\2} \right\} +
\\ 
u_{\pl} \,  \left\{ \exp[ i \, \mbox{\footnotesize $
\sqrt{\frac{2m}{\hbar^{\2}}}$} \, 
\mbox{\footnotesize $\sqrt{V - \sqrt{E^{\2} - |W|^{\2}}  }$}  \, x] \, 
k_{\3} + 
\exp[- \,  i \, \mbox{\footnotesize $
\sqrt{\frac{2m}{\hbar^{\2}}}$} \, 
\mbox{\footnotesize $\sqrt{V - \sqrt{E^{\2} - |W|^{\2}}  }$} 
 \, x] \, k_{\4} \right\}~.
\end{array}
\end{equation}
For $|E| < |W|$,  the solution is given by
\begin{equation}
\label{slos222}
\begin{array}{l}
u_{\mi} \,  \left\{ \exp[   
\mbox{\footnotesize $
\sqrt{\frac{2m}{\hbar^{\2}}  \, \rho}$} 
\, \exp[ i \, \frac{\theta + \pi}{2} ] \, x] \, k_{\1} + 
\exp[- \, \mbox{\footnotesize $
\sqrt{\frac{2m}{\hbar^{\2}}  \, \rho}$} \, 
\exp[ i \, \frac{\theta - \pi}{2} ]
  \, x] \, k_{\2} \right\} +
\\ 
u_{\pl} \,  \left\{ \exp[ \mbox{\footnotesize $
\sqrt{\frac{2m}{\hbar^{\2}} \, \rho}$} \, 
 \exp[ i \, \frac{\pi - \theta }{2} ] 
\, x] \, 
k_{\3} + 
\exp[- \,  \mbox{\footnotesize $
\sqrt{\frac{2m}{\hbar^{\2}}  \, \rho }$} \, 
\exp[ - \, i \, \frac{\theta + \pi}{2}] 
 \, x] \, k_{\4} \right\}~,
\end{array}
\end{equation}
where $\rho = \mbox{\footnotesize $
\sqrt{  V^{\2} + |W|^{\2} - E^{\2}}$}$ and 
$\tan \theta = \mbox{\footnotesize $
\frac{\sqrt{ |W|^{\2} - E^{\2}}}{V}
$}$. In the region of zero potential, by using the boundary conditions
at large distances, 
we find
\begin{equation}
 \Psi(x) = \left\{ \begin{array}{l}
x<0~:  \\
\exp [ \,  
\mbox{\footnotesize $
\sqrt{\frac{2m}{\hbar^{\2}}  \, |E| }$} \,   
x \, ] \, c_{\1} + 
j \, \exp[ \, - \, i \, 
  \mbox{\footnotesize $
\sqrt{\frac{2m}{\hbar^{\2}}  \, |E| }$} \,
x \, ] \, c_{\4}~; \\
 \\
x > a~:  \\
\exp [ \, - \,   
\mbox{\footnotesize $
\sqrt{\frac{2m}{\hbar^{\2}}  \, |E| }$} \,   
x \, ] \, d_{\2} + 
j \, \exp[ \, i \, 
  \mbox{\footnotesize $
\sqrt{\frac{2m}{\hbar^{\2}}  \, |E| }$} \,
x \, ] \, d_{\3}~.
\end{array}  
\right. 
\end{equation}
The matching conditions at the discontinuities of the potential yield
the energy eigenvalues.


\section{CONCLUSIONS}
\label{sec9}

In this paper, we have discussed the resolution of quaternionic, 
$\mathcal{D}_{\Ha} \, \varphi(x) = 0$, 
and complex, $\mathcal{D}_{\Co} \, \varphi(x) = 0$, linear differential 
equations with constant coefficients within a quaternionic formulation of 
quantum mechanics.  We emphasize  that the only {\em quaternionic quadratic} 
equation  involved in the  study of second order linear differential equations
with constant coefficients is given by 
equation~(\ref{qua}) following from $\mathcal{D}_{\Ha} \, \varphi(x) = 0$. 
Due to the right action  of the factor $i$ in complex 
linear differential equations, we cannot factorize a quaternionic 
exponential and consequently we are not 
able to obtain a {\em quaternionic quadratic} equation from 
$\mathcal{D}_{\Co} \, \varphi(x) = 0$.  
Complex linear differential equations can be solved
by searching for quaternionic  solutions of the form $q \, \exp [z \,x ]$, 
where  $q \in \mathbb{H}$ and $z \in \mathbb{C}$. The complex 
exponential factorization gives a  {\em complex quartic} equation.
A similar discussion can be extended to real linear differential equations,
$\mathcal{D}_{\Re} \, \varphi(x) = 0$. In this case, the presence
of left/right operators $\boldsymbol{L}$ and $\boldsymbol{R}$ in 
$\mathcal{D}_{\Re}$ requires quaternionic solutions of the form
$q \, \exp [\lambda \,x ]$, where  $q \in \mathbb{H}$ and 
$\lambda \in \mathbb{R}$. A detailed discussion of real linear differential 
equations deserves a further investigation.

The use of
quaternionic mathematical structures in solving the complex linear
Schr\"odinger equation could represent an important direction for
the search of new physics. The open question whether quaternions could play
a significant role in quantum mechanics is strictly related to the
whole understanding of resolutions of quaternionic differential equations
and eigenvalue problems. The investigation presented in this work is only a 
first step towards a whole theory of quaternionic differential, integral
and functional equations. Obviously, due to the great variety of problems in
using a non-commutative field, it is very difficult to define the precise 
limit  of the subject.


\section*{ACKNOWLEDGEMENTS}

The authors acknowledge the
Department of Physics and INFN, University of Lecce, for the hospitality
and financial support. In particular, they thank 
P.~Rotelli, G.~Scolarici and L.~Solombrino for 
helpful comments and suggestions.


\section*{APPENDIX A: Quaternionic linear quadratic equations}

In this appendix, we give some examples of quaternionic linear quadratic
equations, see cases {\bf (i)}-{\bf (iii)} of subsection \ref{s61}, 
and find their solutions.  

\vspace*{0.3cm}

\noindent
$\bullet$ {\bf (i)}:~~$p^{\2} + \sqrt{2} \, (i+j) \, p - 1 - 2 \,
\sqrt{2} \,  (i+j) = 0$.\\

\noindent In solving such an equation we observe that 
$\boldsymbol{a} = (\mbox{\footnotesize $\sqrt{2}$},
\mbox{\footnotesize $\sqrt{2}$} , 0)$ and $\boldsymbol{c} = - 
(\mbox{\footnotesize $2 \, \sqrt{2}$} , 
\mbox{\footnotesize $2 \, \sqrt{2}$} ,0)$
are parallel vectors,  
$\boldsymbol{c} = - 2 \, \boldsymbol{a}$. Consequently, by
introducing the {\em complex} imaginary unit $\mathcal{I} = (i+j)/\sqrt{2}$, 
we can
reduce the quadratic quaternionic equation to the following 
{\em complex} equation,
\[
p^{\2} + 2 \, \mathcal{I} \, p - 1 - 4 \, \mathcal{I} = 0~,
\]
whose solutions are $p_{\1 , \2} = - \mathcal{I} \pm  2 \, 
\sqrt{\mathcal{I}}$.  It follows that the quaternionic solutions are 
\[
p_{\1 , \2} = \pm  \sqrt{2} - \left( 1 \mp \, \sqrt{2} \right) \, 
\frac{i+j}{\sqrt{2}}~.
\]

\vspace*{0.3cm}

\noindent
$\bullet$ {\bf (ii)}:~~$p^{\2} + i \, p + 
\mbox{$\frac{1}{2}$} \, k = 0$, $\Delta = 0$.\\

\noindent
We note that $\boldsymbol{a} = (1,0,0)$ and  
$\boldsymbol{c} = (0,0, \frac{1}{2})$ are orthogonal vectors and $\Delta =0$.
So, we find two coincident quaternionic solutions given by  
\[
p = - \mbox{$\frac{1}{2}$} \, \boldsymbol{h} \cdot \boldsymbol{a} +  
\boldsymbol{h} \cdot \boldsymbol{a} \times \boldsymbol{c} = 
- \frac{i + j}{2}~.
\]

\vspace*{0.3cm}

\noindent 
$\bullet$ {\bf (ii)}:~~$p^{\2} + j \, p + 1 - k = 0$, 
$\Delta > 0$.\\

\noindent In this case,  
$\boldsymbol{a}=(0,1,0)$ and  $\boldsymbol{c}=(0,0,-1)$ are orthogonal 
vectors, $c_{\0} =1$ and $\Delta = 1/4$. So,
\[
p_{\0} = 0~,~~~x = - \mbox{$\frac{1}{2}$} \pm   \mbox{$\frac{1}{2}$}~,
~~~y=0~,~~~z=1~.
\]
By observing that
\[
\boldsymbol{h} \cdot \boldsymbol{a}=j~,~~~
\boldsymbol{h} \cdot \boldsymbol{c}=-k~,~~~
\boldsymbol{h} \cdot \boldsymbol{a} \times \boldsymbol{c} =-i~,
\]
we find the following quaternionic solutions 
\[ 
p_{\1} = -i~~~~~\mbox{and}~~~~~p_{\2} = - (i + j)~.
\]

\vspace*{0.3cm}

\noindent $\bullet$ {\bf (ii)}:~~$p^{\2} + k \, p + j = 0$, 
$\Delta < 0$.\\

\noindent We have 
$\boldsymbol{a}=(0,0,1)$, $\boldsymbol{c}=(0,1,0)$ and $c_{\0} =0$.
Then $\boldsymbol{a} \cdot \boldsymbol{c}
= 0$ and $\Delta = -3/4$. So,
\[
p_{\0} = \pm   \mbox{$\frac{1}{2}$}~,~~~x = - \mbox{$\frac{1}{2}$}~,
~~~y= \mp \mbox{$\frac{1}{2}$} ~,~~~z=\mbox{$\frac{1}{2}$}~.
\]
In this case,  
\[
\boldsymbol{h} \cdot \boldsymbol{a}=k~,~~~
\boldsymbol{h} \cdot \boldsymbol{c}=j~,~~~
\boldsymbol{h} \cdot \boldsymbol{a} \times \boldsymbol{c} =-i~,
\]
thus, the solutions are given by
\[
p_{\1 , \2} = \mbox{$\frac{1}{2}$} \,  \left( \pm 1 - i \mp j - k \right)~.
\]

\vspace*{0.3cm}

\noindent $\bullet$ {\bf (iii)}:~~$p^{\2} + i \, p + 1 + i + k = 0$.\\

\noindent We have 
$\boldsymbol{a}=(1,0,0)$, $\boldsymbol{c}=(1,0,1)$ and $c_{\0}=1$. 
In this case $\boldsymbol{a} \cdot \boldsymbol{c} \neq 0$, so 
we introduce the quaternion
$ d_{\0} + \boldsymbol{h} \cdot \boldsymbol{d} = 1 + k$, whose vectorial
part $\boldsymbol{d}= \boldsymbol{c} - d_{\0} \boldsymbol{a}= (0,0,1)$
is orthogonal to $\boldsymbol{a}$. The imaginary part of our solution will be 
given in  terms of the imaginary quaternions 
\[
\boldsymbol{h} \cdot \boldsymbol{a}=i~,~~~
\boldsymbol{h} \cdot \boldsymbol{d}=k~,~~~
\boldsymbol{h} \cdot \boldsymbol{a} \times \boldsymbol{d} =-j~.
\]
The real part of $p$ is determined by solving
the equation
\[
16 \, p_{\0}^{\6} + 24 \,  p_{\0}^{\4}  
- 3 \, p_{\0}^{\2} - 1 = 0~.
\]
The real positive solution is given by $p_{\0}^{\2} = \frac{1}{4}$.  
Consequently,
\[
p_{\0} = \pm   \mbox{$\frac{1}{2}$}~,~~~x = - \mbox{$\frac{1}{2}$} \mp 1~,
~~~y= \mp \mbox{$\frac{1}{2}$} ~,~~~z=\mbox{$\frac{1}{2}$}~.
\]
The quaternionic solutions are
\[
p_{\1} = \mbox{$\frac{1}{2}$} \,  \left( 1 - 3 i - j - k \right)~~~~~
\mbox{and}~~~~~
p_{\2} = - \mbox{$\frac{1}{2}$} \,  \left( 1 - i + j - k \right)~.
\]


\section*{APPENDIX B: Quaternionic linear differential equations}

We solve quaternionic linear differential equations whose characteristic
equations are given by the examples {\bf (i)}-{\bf (iii)} in the previous 
appendix.

\vspace*{0.3cm}

\noindent
$\bullet$ {\bf (i)}:~~$\ddot{\varphi}(x) + \sqrt{2} \, ( \, i+j \, ) \, 
\dot{\varphi}(x) - \left[ \, 1 + 2 \sqrt{2} \, ( \, i+j \, ) \, \right] \,
\varphi(x) = 0~,~~~\varphi(0)=i~,~~~
\dot{\varphi}(0)= \mbox{$\frac{1+k}{\sqrt{2}}$}$~.\\

\noindent 
The exponential $\exp[ \, p \, x \, ]$ is solution of the previous 
differential equation if and only if the quaternion $p$ satisfies the 
following quadratic equation
\[
p^{\2} + \sqrt{2} \, (i+j) \, p - 1 - 2 \,
\sqrt{2} \,  (i+j) = 0~,\]
whose solutions are given by 
\[
p_{\1 , \2} = \pm  \sqrt{2} - \left( 1 \mp \, \sqrt{2} \right) \, 
\frac{i+j}{\sqrt{2}}~.
\]
Consequently,
\begin{eqnarray*}
\varphi(x) &  =  &
\exp \left\{ \left[ \, 
 \sqrt{2} - \left( 1 - \sqrt{2} \right) \,  
\frac{i+j}{\sqrt{2}} \,  \right] \, x \, \right\} \, c_{\1}
+
\exp \left\{ \left[ \, 
- \, \sqrt{2} - \left( 1 +  \sqrt{2} \right) \,  
\frac{i+j}{\sqrt{2}} \, \right] \, x \, \right\} \, c_{\2}~.
\end{eqnarray*}
By using the initial conditions, we find
\[
\varphi(x) = 
\exp 
\left[ \, - \, \frac{i+j}{\sqrt{2}} \, x \, \right] \, 
\cosh \left[ \left( \,
\sqrt{2} + \frac{i+j}{\sqrt{2}} \, \right) \, x \, \right] \, i~.
\]

\vspace*{0.3cm}

\noindent
$\bullet$ {\bf (ii)}:~~$\ddot{\varphi}(x) + (\, 1+i \, ) \, 
\dot{\varphi}(x) + \frac{2 + i + k}{4} \,
\varphi(x) = 0~,~~~\varphi(0)=0~,~~~
\dot{\varphi}(0)= \, - \, \mbox{$\frac{1+i+j}{2}$}$~.\\

\noindent We look for exponential solutions of the form  
$\varphi(x) = \exp [ \, q \, x \, ] = 
\exp [ \, ( \, p - \frac{1}{2} \, )\, x \, ]$. The quaternion $p$
must satisfy the quadratic equation
\[ p^{\2} + i \, p + 
\mbox{$\frac{1}{2}$} \, k = 0~.\]
This equation implies 
\[
p_{\1}= p_{\2} =   
- \, \frac{i + j}{2}~.
\]
Thus,
\[
\varphi_{\1}(x) = \exp 
\left[ \, - \, \frac{1 + i + j}{2} \, x \, \right]~.
\]
The second linearly independent solution is given by
\[
\varphi_{\2}(x) = \left( \, x + i \, \right)\exp 
\left[ \, - \, \frac{1 + i + j}{2} \, x \, \right]~.
\]
By using the initial conditions, we find
\[
\varphi(x) = \left\{ \, \exp [ \, q \, x\, ] + 
( \, x + i \, ) \,   \exp [ \, q \, x \, ] \, i \, \right\} \, 
\left[ \, 1 + q^{\mi \1} \, ( \, 1 + i \, q \, ) \, i \, \right]^{\mi \1}~,
\]
where $q =   \, - \, ( \, 1 + i + j \, ) \, / \, 2$.

\vspace*{0.3cm}

\noindent 
$\bullet$ {\bf (ii)}:~~$\ddot{\varphi}(x) + ( \, 2+j \, ) \, 
\dot{\varphi}(x) + \left( \, 2 + j - k \, \right) \,
\varphi(x) = 0~,~~~\varphi(0)= \, \mbox{$\frac{1-i}{2}$}~,~~~
\dot{\varphi}(0)= j$~.\\

\noindent The exponential solution $\varphi(x) = \exp [ \, q \, x \, ] = 
\exp [ \, ( \, p - 1 \, )\, x \, ]$ leads to
\[ p^{\2} + j \, p + 1 - k = 0~, 
\]
whose solutions are
\[ 
p_{\1} = -i~~~~~\mbox{and}~~~~~p_{\2} = - (i + j)~.
\]   
Consequently,
\[
\varphi(x) = \exp[ \, - \, x \, ] \,
\left\{ \, \exp [\, - \, i \, x \, ] \, c_{\1} + 
 \exp [\, - \, ( \, i + j \, )  \, x \, ] \, c_{\2} \, \right\}~.
\]
The initial conditions yield 
\[
 \varphi(x) = \exp[ \, - \, x \, ] \,
\left\{ \, \exp [\, - \, i \, x \, ] \, \mbox{$\frac{3 - i - 2j}{2}$} + 
 \exp [\, - \, ( \, i + j \, )  \, x \, ] \, ( \, j - 1 \, ) \, \right\}~.
\]

\vspace*{0.3cm}

\noindent $\bullet$ {\bf (ii)}:~~$\ddot{\varphi}(x) + k \, 
\dot{\varphi}(x) + j \, 
\varphi(x) = 0~,~~~\varphi(0)= i + k ~,~~~
\dot{\varphi}(0)= 1$~.\\

\noindent The characteristic equation is
\[ p^{\2} + k \, p + j = 0~,\]
whose solutions are
\[
p_{\1 , \2} = \mbox{$\frac{1}{2}$} \,  \left( \pm 1 - i \mp j - k \right)~.
\]
Thus, the general solution of our differential equation reads
\[
\varphi(x) = 
\exp [\, \mbox{$\frac{1-i-j-k}{2}$} \, x \, ] \, c_{\1} + 
\exp [\, - \, \mbox{$\frac{1+i-j+k}{2}$} \, x \, ] \, c_{\2}~.
\]
By using the initial conditions, we obtain
\[
\varphi(x) = \left\{ \, 
\exp [\, \mbox{$\frac{1-i-j-k}{2}$} \, x \, ] + 
\exp [\, - \, \mbox{$\frac{1+i-j+k}{2}$} \, x \, ] \, \right\} \, 
\mbox{$\frac{i+k}{2}$}~.
\]

\vspace*{0.3cm}

\noindent $\bullet$ {\bf (iii)}:~~
$\ddot{\varphi}(x) + \left( \, i - 2 \, \right) \, 
\dot{\varphi}(x) + \left( \, 2 + k \, \right) \, 
\varphi(x) = 0~,~~~\varphi(0)= 0 ~,~~~
\dot{\varphi}(0)= j$~.\\

\noindent 
By substituting $\varphi(x) = \exp [ \, q \, x \, ] = 
\exp [ \, ( \, p + 1 \, )\, x \, ]$ in the previous
differential equation, we find 
\[p^{\2} + i \, p + 1 + i + k = 0~.\]
The solutions of this quadratic quaternionic equation are
\[
p_{\1} = \mbox{$\frac{1}{2}$} \,  \left( 1 - 3 i - j - k \right)~~~~~
\mbox{and}~~~~~
p_{\2} = - \mbox{$\frac{1}{2}$} \,  \left( 1 - i + j - k \right)~.
\]
So, the general solution of the differential equation is
\[
\varphi(x) = 
\exp [\, \mbox{$\frac{1-3i-j-k}{2}$} \, x \, ] \, c_{\1} + 
\exp [\, - \, \mbox{$\frac{1-i+j-k}{2}$} \, x \, ] \, c_{\2}~.
\]
By using the initial conditions, we obtain
\[
\varphi(x) = \left\{ \, 
\exp [\, \mbox{$\frac{1-3i-j-k}{2}$} \, x \, ] - 
\exp [\, - \, \mbox{$\frac{1-i+j-k}{2}$} \, x \, ] \, \right\} \, 
\mbox{$\frac{j-i+2k}{6}$}~.
\]


\section*{APPENDIX C: Diagonalization and Jordan form}

In this appendix, we find the  solution of quaternionic and complex linear
differential equations by using diagonalization and Jordan form.

\subsection*{Quaternionic linear differential equation}

By using the discussion about quaternionic quadratic equation,
it can immediately be shown that the solution of the 
following second order equation
\begin{equation*}
\ddot{\varphi}(x) + ( \, k - i \, ) \, \dot{\varphi}(x) 
- j \, \varphi(x) = 0~,
\end{equation*}
with initial conditions
\[ \varphi(0) = \mbox{$\frac{k}{2}$}~,~~~\dot{\varphi}(0)=1 + 
\mbox{$\frac{j}{2}$}~,\]
is given by
\begin{equation*}
\varphi(x) = \left( \, x + \mbox{$\frac{k}{2}$} \, \right) \, 
\exp[ \, i \, x ]~.
\end{equation*} 
Let us solve this differential equation by using its matrix form 
(\ref{matrix}), with
\[
M = \left( \begin{array}{cc} 0 & ~1 \\
j  & ~i-k 
\end{array} \right)~.
\]
This quaternionic matrix can be reduced to its Jordan form
\[
M = J \, \left( \begin{array}{cc} i & ~1 \\
0 &  ~i
\end{array} \right)
\, J^{\mi \1}~.
\]
by the matrix transformation 
\[
J  =  \left( \begin{array}{cc} 
1 & ~\mbox{$ \frac{k}{2}$} \\
i & ~1+ \mbox{$ \frac{j}{2}$} 
\end{array} \right)
~,~~~
J^{\mi \1}  = \left( \begin{array}{cc} 
\mbox{$ \frac{3+j}{4}$}  & 
~$-$ \mbox{$ \frac{i+k}{4}$} \\
$-$ \mbox{$ \frac{i+k}{2}$}  & 
~\mbox{$ \frac{1-j}{2}$} 
\end{array} \right)
\]
\noindent 
The solution  of the quaternionic linear quaternionic differential 
equation  is then given by
\begin{eqnarray*}
\varphi (x) & = & J_{\1 \1} \, 
\exp \left[ \, i \, x \, \right] \, \left[  
J_{\1 \1}^{\mi \1} \, \, \varphi(0) + 
J_{\1 \2}^{\mi \1} \, \, \dot{\varphi}(0) \right] + \nonumber \\
 & &  
\left( \, x \, J_{\1 \1} + J_{\1 \2}  \, \right) \, 
\exp \left[ \,  i \, x \, \right]
 \, \left[
J_{\2 \1}^{\mi \1} \, \, \varphi(0) +
J_{\2 \2}^{\mi \1} \, \, \dot{\varphi}(0) \right]
\nonumber \\
 & = & \left( \, x \, J_{\1 \1} + J_{\1 \2}  \, \right) \, 
\exp \left[ \,  i \, x \, \right] \nonumber \\
& = & 
\left( \, x + \mbox{$\frac{k}{2}$} \, \right) \, 
\exp[ \, i \, x ]~. 
\end{eqnarray*}

\subsection*{Complex linear differential equations}

Let us now consider the complex linear quaternionic differential equation
\begin{equation*}
\ddot{\varphi}(x) - j \, \varphi(x) \, i = 0~,
\end{equation*}
with initial conditions
\[ \varphi(0) = j~,~~~\dot{\varphi}(0)=k~.\]
To find particular solutions,
we set $\varphi(x) = q \, \exp [ \, z \, x ]$.
Consequently, 
\[ q \, z^{\2} - j \, q \, i = 0~.\]
The solution of the complex linear second order differential equation
is
\begin{equation*}
\varphi(x) = \mbox{$\frac{1}{2}$} \, \left[
(i+j) \, \exp[-ix] + (j-i) \, \cosh x + (k -1) \, \sinh x
\right]~.
\end{equation*} 
This solution can also be obtained by using the matrix 
\[
M_{\Co} = \left( \begin{array}{cc} 0 & ~1 \\
j \, R_{\i} & ~0 
\end{array} \right)~,
\]
and its diagonal form
\[
M_{\Co} = S_{\Co} \, \left( \begin{array}{cc} - i \, R_{\i} & ~0 \\
0 &  ~i 
\end{array} \right)
\, S_{\Co}^{\mi \1}~,
\]
where
\[
S_{\Co}  =  \left( \begin{array}{cc} 
\mbox{\footnotesize $ \frac{1-i-j-k}{2}$} + 
\mbox{\footnotesize $ \frac{1-i+j+k}{2}$} \, \mbox{\footnotesize $R_{\i}$}
 & 
~~ \, \, \mbox{\footnotesize $ \frac{1+i-j+k}{2}$} - 
\mbox{\footnotesize $ \frac{1+i+j-k}{2}$} \, \mbox{\footnotesize $R_{\i}$}
 \\
\mbox{\footnotesize $ \frac{1+i+j-k}{2}$} - 
\mbox{\footnotesize $ \frac{1+i-j+k}{2}$} \, \mbox{\footnotesize $R_{\i}$}
 & 
~\mbox{\footnotesize $ - \frac{1-i-j-k}{2}$} + 
\mbox{\footnotesize $ \frac{1-i+j+k}{2}$} \, \mbox{\footnotesize $R_{\i}$}
\end{array} \right)
\]
and
\[
S_{\Co}^{\mi \1}  = \mbox{$\frac{1}{4}$} \left( \begin{array}{cc} 
\mbox{\footnotesize $ \frac{1+i+j+k}{2}$} - 
\mbox{\footnotesize $ \frac{1+i-j-k}{2}$} \, \mbox{\footnotesize $R_{\i}$}
 & 
~~ \, \,
\mbox{\footnotesize $ \frac{1-i-j+k}{2}$} + 
\mbox{\footnotesize $ \frac{1-i+j-k}{2}$} \, \mbox{\footnotesize $R_{\i}$}
 \\
\mbox{\footnotesize $ \frac{1-i+j-k}{2}$} + 
\mbox{\footnotesize $ \frac{1-i-j+k}{2}$} \, \mbox{\footnotesize $R_{\i}$}
 & 
~\mbox{\footnotesize $ - \frac{1+i+j+k}{2}$} - 
\mbox{\footnotesize $ \frac{1+i-j-k}{2}$} \, \mbox{\footnotesize $R_{\i}$}
\end{array} \right)~.
\]
\noindent 
The solution  of the complex linear quaternionic differential equation  is
then given by
\begin{eqnarray*}
\varphi (x) & = & S_{\Co \, \1 \1} \, 
\exp \left[ \, -  \, i \, R_{\i}
\, x \, \right] \, \left[  
S_{\Co \, \1 \1}^{\mi \1} \, \, \varphi(0) + 
S_{\Co \, \1 \2}^{\mi \1} \, \, \dot{\varphi}(0) \right] + \nonumber \\
 & &  
S_{\Co \, \1 \2} \, \exp \left[ \,  i \, x \, \right]
 \, \left[
S_{\Co \, \2 \1}^{\mi \1} \, \, \varphi(0) +
S_{\Co \, \2 \2}^{\mi \1} \, \, \dot{\varphi}(0) \right]
\nonumber \\
 & = & \mbox{$\frac{1}{4}$} \, \left\{ ( 1 - i + j - k) \, \exp [-x] -
( 1 + i - j - k) \, \exp [x] \right\} + 
\nonumber \\
 & &   \mbox{$\frac{i + j }{2}$} \, \exp [-ix]~. 
\end{eqnarray*}



\end{document}